\documentclass[a4paper,fleqn,usenatbib,useAMS]{mnras}

\usepackage{graphicx}	
\usepackage{amsmath}	
\usepackage{amssymb}	
\usepackage{multicol}        
\usepackage{bm}		
\usepackage{pdflscape}	
\usepackage{multirow}
\usepackage{xcolor}
\usepackage{hyperref}
\usepackage{booktabs}

\def\specialname[#1]{\textbf{\textsc{#1}}}

\newcommand{\tabincell}[2]{\begin{tabular}{@{}#1@{}}#2\end{tabular}}
\usepackage{microtype}

\usepackage[T1]{fontenc}
\usepackage{ae,aecompl}

\usepackage{newtxtext,newtxmath}

\title[Finding proto-clusters to trace galaxy evolution]{Finding proto-clusters
to trace galaxy evolution: I. The finder and
its performance}

\author[Wang et al.]{
	Kai Wang,$^{1,2}$\thanks{Contact e-mail: wkcosmology@gmail.com\href{}{}}\thanks{Present address: Department of Astronomy, University of Massachusetts Amherst, MA 01003, USA}
	H.J. Mo,$^{2}$
	Cheng Li,$^{1}$
	and Yangyao Chen$^{1,2}$
	\\
	$^{1}$Department of Astronomy, Tsinghua University, Beijing 100084, China\\
	$^{2}$Department of Astronomy, University of Massachusetts Amherst, MA 01003, USA}

\date{Last updated 2020 May 22; in original form 2018 September 5}

\pubyear{2020}

\begin{document}
	\label{firstpage}
	\pagerange{\pageref{firstpage}--\pageref{lastpage}}
	\maketitle


\begin{abstract}
    We develop a method to identify proto-clusters based on dark matter halos
    represented by galaxy groups selected from surveys of galaxies at high
    redshift. We test the performance of this method on halos in N-body simulations,
    and find that it can correctly identify more than $85\%$ of the true
    proto-clusters with $\gtrsim 95\%$ purity and with mass estimates typically
    within $0.25$ dex from their true values. We show how the information
    provided by the proto-clusters can be used to link galaxies in present-day
    clusters of galaxies with their high redshift progenitors. Our tests show
    that the proto-clusters identified by our method can recover reliably the
    progenitor stellar mass distribution of galaxies, thereby providing an
    avenue to investigate the formation and evolution of present-day galaxy
    clusters and their member galaxies.
\end{abstract}





\section{Introduction}%
\label{sec:introduction}

Large surveys of galaxies are now available to investigate the statistical
properties of the galaxy population at different redshifts. These observations
can be used to study the evolution of the galaxy population over the history of
the Universe. However, since galaxies at different redshifts have no individual
causal relations, one can only study the evolution in a statistical way, by
comparing the distribution functions of galaxy properties, such as stellar
mass, star formation rate and spatial clustering. In order to use the
observational data to make accurate and unbiased inferences on the evolution,
it is necessary to connect galaxy populations at different redshifts in a
meaningful way.

In the literature, two approaches have been adopted to connect galaxies at
different redshift. The first is based on abundance matching, which assumes
that ranks of galaxies in stellar mass are preserved as galaxies evolve with
redshift \citep[e.g.][]{vandokkumGROWTHMASSIVEGALAXIES2010,
    behrooziUSINGCUMULATIVENUMBER2013, torreyAnalysisEvolvingComoving2015,
    torreyForwardBackwardGalaxy2017, hillMassColorStructural2017,
mendelKinematicsMassiveQuiescent2020}. Since this method uses only stellar mass
to establish the connection, it ignores information about the evolution carried
by other properties, such as the stellar population and environment of galaxies
\citep{clauwensLargeDifferenceProgenitor2016}. More recently,
\cite{kipperRoleStochasticSmooth2021} proposed a model for the evolution of the
stellar mass and star formation rate of galaxies, and constrained the model
parameters using observational data. However, this method is model-dependent
and may be inaccurate when the redshift range covered is large. In addition,
since galaxy properties are known to depend on environment, such as their halos
\citep[e.g.][]{weinmannPropertiesGalaxyGroups2006, wangELUCIDIVGalaxy2018},
ignoring the environmental difference of galaxies will mix different
evolutionary tracks, making the results difficult to interpret.

In the current paradigm of structure formation, galaxies form and evolve in
dark matter halos, which are the virialized parts of the cosmic structure
formed hierarchically through gravitational instability. Thus, the properties
of the host halos of galaxies can provide additional information about the
evolutionary tracks of galaxies. The formation and evolution of the halo
population can be well understood using the state-of-the-art N-body
simulations, and the link of a halo to its progenitors is provided by its
merging tree. Since the progenitors of a galaxy hosted by a halo at the present
day must be the galaxies that have formed in the progenitor halos, a link
between the galaxy and its progenitors can be made through the connection of
the halo with its progenitors. The second approach to connect galaxies across
different redshifts is, therefore, to identify the progenitors of the most
massive cluster galaxies in the local Universe as the most massive galaxies in
high-$z$ halos that are expected to evolve to $z=0$ halos of mass similar to
that of the clusters in question
\citep[e.g.][]{lidmanEvidenceSignificantGrowth2012, cookeStellarMassGrowth2019,
demaioGrowthBrightestCluster2020, linSTELLARMASSGROWTH2013}. This method is
valid only if mergers among the high-$z$ halos are negligible in the subsequent
evolution. To overcome this problem,
\cite{zhaoExploringProgenitorsBrightest2017} developed a hybrid method to link
the brightest cluster galaxies at low $z$ to the brightest galaxies  in the
high density regions at high $z$, with the size of each of the regions chosen
large enough so that subsequent mergers among the regions are negligible.    
However, since these investigations only traced the evolution of the  most
massive cluster galaxies following the main branches of the halo  merger trees,
they ignored a large number of galaxies that will evolve into satellite
galaxies in $z=0$ clusters. Clearly, a more general method using the
information provided by the whole halo merger tree is needed to connect all
galaxies in present-day clusters to their high-$z$ progenitors.

The main objective of this paper is to develop a method that can link cluster
galaxies to their progenitors reliably. A key component in our method is to
group galaxies at high-$z$ into common halos and identify proto-clusters that
will evolve into clusters of given mass at the present time. As shown in
\citet{wangIdentifyingGalaxyGroups2020}, with the high-$z$ surveys of galaxies
available now and in the near future, one can identify reliably galaxy
groups/clusters to represent dark matter halos over a large mass range. One
focus of the present paper is to develop a method to identify proto-clusters
from such surveys. A number of proto-cluster identification methods have been
proposed in the literature \citep[See][for a
review]{overzierRealmGalaxyProtoclusters2016}, using densities defined by
normal galaxies \citep{chiangANCIENTLIGHTYOUNG2013,
    chiangDISCOVERYLARGENUMBER2014, dienerPROTOGROUPSZCOSMOSDEEPSAMPLE2013,
    franckCANDIDATECLUSTERPROTOCLUSTER2016,
    toshikawaSYSTEMATICSURVEYPROTOCLUSTERS2016,
lovellCharacterisingIdentifyingGalaxy2018}, Ly-$\alpha$ emitters
\citep{chiangSURVEYINGGALAXYPROTOCLUSTERS2015a}, Ly-$\alpha$ absorption systems
\citep{caiMAPPINGMOSTMASSIVE2016, leeSHADOWCOLOSSUS442016,
caiMappingMostMassive2017}, and star formation rate
\citep{martinacheSpitzerPlanckHerschel2018}. All these methods have to be
calibrated using semi-analytical models and/or hydrodynamic simulations to
ensure that the identified systems represent proto-clusters with well-defined
mass. In comparison, our method based on halos mitigates the uncertainties
introduced by baryonic processes in galaxy formation, so that it can be tested
and calibrated using cosmological N-body simulations.

The paper is organized as follows. The simulation data and the empirical model
of galaxy formation used for our analyses are presented in \S\,\ref{sec:data}.
Our proto-cluster finder and the test results of its performances are presented
in \S\,\ref{sec:finder}. We describe how to use the information provided by
proto-clusters to link halos and galaxies at different redshifts in
\S\,\ref{sec:linking_high_z_progenitors_to_local_clusters}. Finally, we
summarize our main results in \S\,\ref{sec:summary}.

\section{Simulation Data for Testing}%
\label{sec:data}

\begin{table}
	\centering
	\caption{Number of halos in the N-body simulation.}
	\label{tab:halo_statistics}
	\begin{tabular}{ccccc} 
		\toprule
        $\log \left(M_h/[h^{-1}M_{\odot}]\right)$ & $z=0$   & $z=1$   & $z=2$   \\
         \midrule
        $[12.0, ~~~\infty~]$       & 397,850 & 329,127 & 172,112 \\
         \midrule
        $[14.0, 14.2]$              & 1,307   & 161     & 5       \\
         \midrule
        $[14.2, 14.5]$              & 787     & 54      & 0       \\
         \midrule
        $[14.5, 15.0]$              & 257     & 4       & 0       \\
         \bottomrule
	\end{tabular}
\end{table}

\subsection{The simulation and the empirical model of halo occupation}%
\label{sub:the_simulation_and_empirical_model}

We use the cosmological simulation, ELUCID
\citep{wangELUCIDEXPLORINGLOCAL2016}, combined with an empirical model of
galaxy formation to construct mock galaxy samples to test our method. ELUCID
was run with L-GADGET, a memory-optimized version of GADGET-2
\citep{springelSimulationsFormationEvolution2005}, using $3072^3$ dark matter
particles, each with a mass of $3.09\times 10^{8} h^{-1}M_{\odot}$, in a
periodic box with a side length of $500h^{-1}\rm cMpc$. The simulation uses
cosmological parameters based on WMAP5
\citep{dunkleyFIVEYEARWILKINSONMICROWAVE2009}: $\Omega_{\rm m}=0.258$,
$\Omega_{\Lambda} =0.742$, $H=100h~\rm km~s^{-1}~Mpc^{-1}$ with $h=0.72$, and
$\sigma_8=0.80$. The simulation covers the structure evolution from $z=100$ to
$0$, and records 100 snapshots from $z=19$ to $0$. Dark matter halos and
subhalos are identified using the friend-of-friend (FoF) and SUBFIND algorithms
\citep{springelPopulatingClusterGalaxies2001}, and halo merger trees are
constructed to trace the merging histories of individual halos using the code
provided by \citet{springelSimulationsFormationEvolution2005}.

We populate dark matter halos in ELUCID with galaxies using the empirical model
developed in \citet{luEmpiricalModelStar2014, luStarFormationStellar2015}. This
model treats central and satellite galaxies separately. For central galaxies,
the star formation rate is parameterized as a function of redshift and host
halo mass. For satellite galaxies, the star formation rate is assumed to
decline with time until the satellite merges with the central galaxy. The free
parameters are constrained with the observed galaxy stellar mass function
spanning a large range of redshift and the cluster galaxy luminosity function
in the low-$z$ Universe. The positions and velocities of individual galaxies
are assigned according to those of halos (for central galaxies) and subhalos
(for satellite galaxies). The details of the implementation of the empirical
model to the simulation can be found in \citet{chenELUCIDVICosmic2019}.

\subsection{Proto-clusters in the simulation}%
\label{sub:proto_clusters_in_the_simulation}

We use all dark matter halos with mass above $10^{12}h^{-1}M_{\odot}$ at a
given redshift $z$ to trace proto-clusters. This choice of mass threshold is
motivated by the fact that such halos can be identified reliably as galaxy
groups at different redshifts \citep{wangIdentifyingGalaxyGroups2020,
looserOptimizingHighRedshift2021}. In this paper, we use halos at $z=0$, $1$
and $2$ (see Table\,\ref{tab:halo_statistics}). We define a proto-cluster as
the set of dark matter halos at $z>0$ that end up in a common descendant halo
at $z=0$. Thus, a proto-cluster is the collection of progenitor halos at $z>0$
for a dark matter halo at $z=0$. In the literature, investigations of
proto-clusters have been focused on relatively massive systems which correspond
to massive dark matter halos at $z=0$, e.g.\ with halo mass $M_0\gtrsim
10^{14}h^{-1}M_{\odot}$. Thus, some of the massive progenitors contained in a
proto-cluster may themselves be high-$z$ clusters according to our definition.
For clarity, we use $M_0$ to denote the descendant halo mass at $z=0$, and
$M_h$ to denote the progenitor dark matter halo mass (See
Table\,\ref{tab:terminology}). We retrieve all the progenitor halos with $M_h
\geq 10^{12}h^{-1}M_{\odot}$ for each descendant halo with $M_0\geq
10^{12}h^{-1}M_{\odot}$ following its merger tree in the simulation. We thus
obtain a set of samples of present-day halos and their true proto-clusters at a
given high redshift. These links between present-day halos and their
proto-clusters will be used to calibrate and test our method to identify
proto-clusters from observational data.

\subsection{Redshift-space distortion}%
\label{sub:redshift_space_distortion}

In real observations, we can only infer the position of a galaxy from $({\rm
RA,~Dec,~}z)$, where the first two specify the position of an object in the
sky, while the redshift $z$ can be converted to a line-of-sight distance.
However, the distance obtained from $z$ is contaminated by the peculiar
velocity of the object owing to the redshift-space distortion. For our problem,
this contamination can be divided into two categories: small-scale
Finger-of-God effect caused by the virial motion of galaxies inside individual
dark matter halos \citep[see][]{jacksonCritiqueReesTheory1972}, and large-scale
Kaiser effect owing to the peculiar motion of dark matter halos produced by the
gravitational interactions on super-halo scales
\citep[see][]{kaiserClusteringRealSpace1987}. Since our method relies on halos
to trace proto-clusters, and since the Finger-of-God effect is corrected in the
halo/group finding process \citep{yangHalobasedGalaxyGroup2005}, we only need
to consider the Kaiser effect.

To mimic the Kaiser effect in our analysis, we modify the positions of dark
matter halos along one chosen direction, assumed to be the $\boldsymbol{\hat X}$
direction. For a dark matter halo at
$\boldsymbol{r}=X\boldsymbol{\hat X} + Y\boldsymbol{\hat Y} +
Z\boldsymbol{\hat Z}$
with peculiar velocity $\boldsymbol{v}$ in a box with redshift $z$,
we update its $\boldsymbol{\hat X}$-component by
\begin{equation}
    X \leftarrow X - D_c(z) + D_c(z^{\prime}), \quad
    z^{\prime} \equiv (z + 1)\left(\frac{\boldsymbol{v}\cdot \boldsymbol{\hat X}}{c}
    + 1\right) - 1
\end{equation}
where $D_c(z)$ is the comoving distance at redshift $z$, and $c$ is the speed
of light. Halos near the edge of the simulation box are properly taken care of
by using the periodic boundary conditions.

\subsection{Descendant halo mass calibration}%
\label{sub:descendant_halo_mass_calibration}
\begin{figure*}
    \centering
    \includegraphics[width=0.8\linewidth]{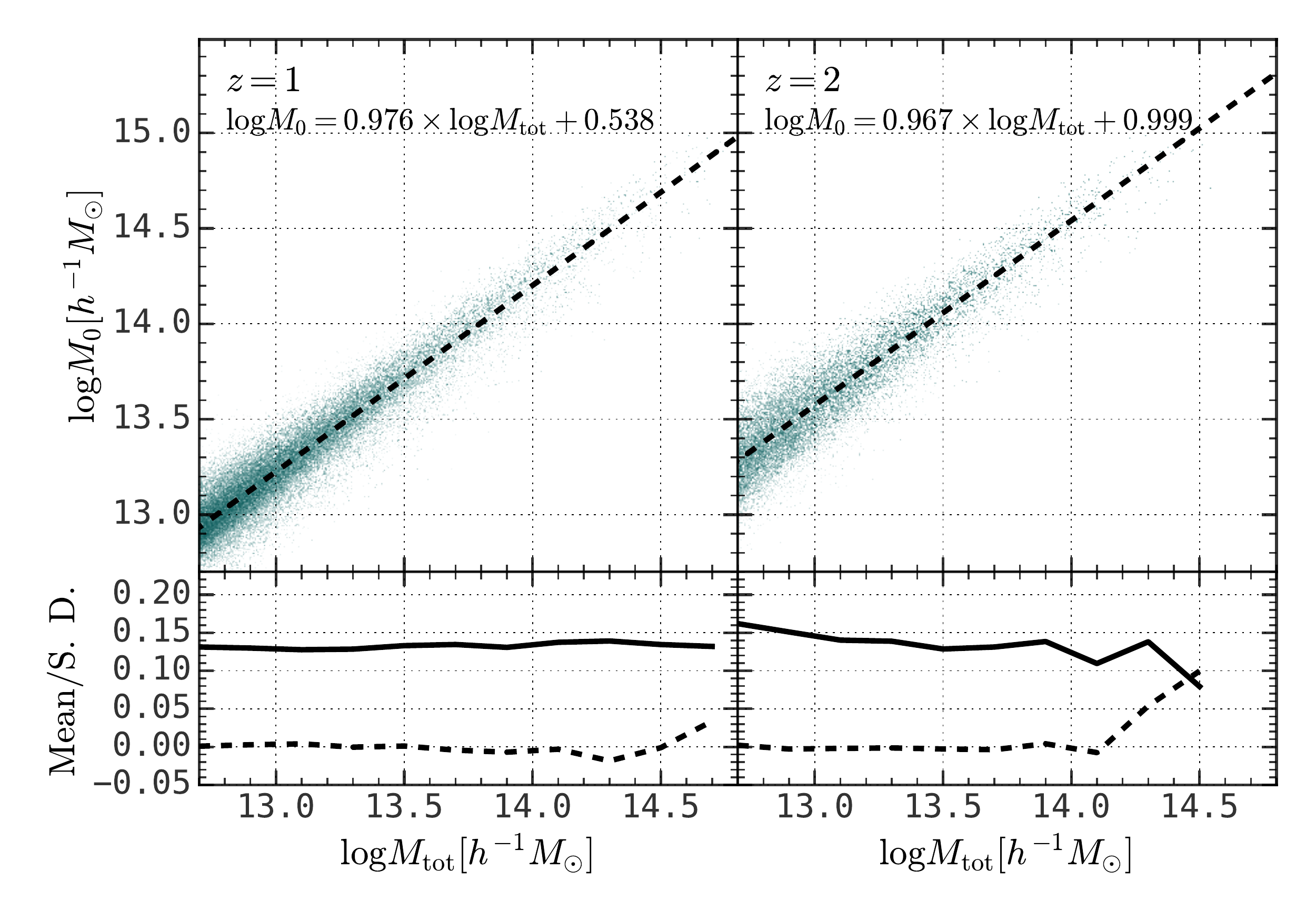}
    \caption{The \textit{upper} panels show the relation between the total halo
        mass, $M_{\rm tot}$, and the descendant halo mass, $M_0$, and the lines are
        the fitting functions shown in the upper left corner. In the
        \textit{lower} panels, the solid lines show the standard deviation from
        the fitting function, and the dashed lines show the mean of $\log(M_{0,
        \rm fitting} / M_0)$, where $\log(M_{0, \rm fitting})$ is the
        descendant mass obtained from the fitting function. The \textit{left}
        panels are for $z=1$, and the \textit{right} panels for $z=2$.
    }%
    \label{fig:figure/halo_masss_z0_cali}
\end{figure*}

Descendant halo mass plays a critical role in connecting galaxies at high-$z$
to their low-$z$ counterparts. In the literature the descendant halo mass is
usually estimated from the overdensity of a given class of tracers, such as
galaxies, halos and dark matter \citep[e.g.][]{chiangANCIENTLIGHTYOUNG2013,
    steidelLargeStructureGalaxies1998,
steidelSpectroscopicIdentificationProtocluster2005}. Here we choose to
calibrate the descendant halo mass using the total halo mass, $M_{\rm tot}$,
which is the sum of the masses of all the halos in the proto-cluster with
$M_h\geq 10^{12}h^{-1}M_{\odot}$. As shown in
Fig.\,\ref{fig:figure/halo_masss_z0_cali}, there is a well-defined relation
between $M_{\rm tot}$ and $M_0$, which is well described by a linear function.
The standard deviation of the relation is quite small, typically below 0.15 dex
for the descendant mass $M_0>10^{13}h^{-1} M_\odot$. Thus, the masses of the
proto-clusters can be estimated reliably using calibrations from $N$-body
simulations. We will use such calibrations to estimate the descendant halo
masses for candidate proto-clusters.

\section{The proto-cluster finding algorithm and its performance}
\label{sec:finder}

In this section, we present a proto-cluster finding algorithm based on the
Friends-of-Friends (FoF) method, and test its performance. The FoF algorithm
was used to identify dark matter halos in N-body simulations and to identify
galaxy groups in galaxy surveys \citep{davisEvolutionLargescaleStructure1985,
    ekeGalaxyGroups2dFGRS2004, knobelOPTICALGROUPCATALOG2009,
wangIdentifyingGalaxyGroups2020}. Our proto-cluster finding algorithm uses the
distribution of dark matter halos in redshift space, where halos are assumed to
be identified through a group finding process, such as those described in
\citet{yangHalobasedGalaxyGroup2005, yangGalaxyGroupsSDSS2007} and
\citet{wangIdentifyingGalaxyGroups2020}. This approach mitigates all the
baryon-related physics in identifying virialized halos and is, therefore,
applicable as long as a complete sample of relatively massive halos/groups
(e.g. $M_h> 10^{12}h^{-1}M_{\odot}$) is available.

\subsection{The proto-cluster finder}%
\label{sub:fof}

For a given dark matter halo distribution in the redshift space, we group the
$i$-th and $j$-th halos together if they satisfy the following criteria:
\begin{align}
    &l_{\parallel}\cdot R_{\rm vir} \geq |X_i - X_j| \\
    &l_{\perp} \cdot R_{\rm vir} \geq \sqrt{(Y_j - Y_k)^2 + (Z_j - Z_k)^2}
\end{align}
where $l_{\parallel}$ and $l_{\perp}$ are two free parameters, $R_{\rm vir} =
\max(R_{{\rm vir}, i},~ R_{{\rm vir}, j})$, and the halo virial radii, $R_{{\rm
vir}, i}$ and $R_{{\rm vir}, j}$, are calculated using the package
\texttt{Halotools} \citep{hearinForwardModelingLargescale2017}. Again, $X$
is the distance along the line-of-sight, while $Y$ and $Z$ are in the
perpendicular directions.

For all halos that are linked into a candidate proto-cluster according to the
above criteria, we calculate a total mass, $M_{\rm tot}$, which is the sum of
the masses of all these halos. We then assign to each candidate proto-cluster a
descendant halo mass, $M_{\rm 0, e}$, using the fitting function in
Fig.\,\ref{fig:figure/halo_masss_z0_cali}.

\begin{table*}
    \centering
    \caption{Terminologies and symbols used in the paper.}
    \label{tab:terminology}
    \begin{tabular}{ll}
     \toprule
    \tabincell{l}{\bf Terminology}                                 & \tabincell{l}{\bf Explanation}                                       \\
     \midrule
    \tabincell{l}{true proto-cluster}                              & \tabincell{l}{the collection of the progenitor halos and galaxies for    \\ a dark matter halo at $z=0$} \\
     \midrule
    \tabincell{l}{candidate proto-cluster}                         & \tabincell{l}{the collection of halos  identified by the proto-cluster finder} \\
     \midrule
    \tabincell{l}{halo mass ($M_h$)}                               & \tabincell{l}{mass of a dark matter halo}                          \\
     \midrule
    \tabincell{l}{descendant halo mass ($M_{0}$)}                  & \tabincell{l}{mass of the descendant dark matter halo at $z=0$}              \\
    \midrule
    \tabincell{l}{estimated descendant halo mass ($M_{\rm 0, e}$)} & \tabincell{l}{estimated mass of the descendant dark matter halo at $z=0$}              \\
    \midrule
    \tabincell{l}{stellar mass of galaxy ($M_*$)}                            & \tabincell{l}{stellar mass of a galaxy}                                          \\
     \midrule
    \tabincell{l}{stellar mass of descendant galaxy ($M_{*,0}$)}                        & \tabincell{l}{stellar mass of a descendant galaxy at $z=0$}            \\
     \bottomrule
    \end{tabular}
\end{table*}

\subsection{Completeness and purity of the identified proto-cluster population}%
\label{ssub:group_level_performance}

\begin{figure*}
    \centering
    \includegraphics[width=\linewidth]{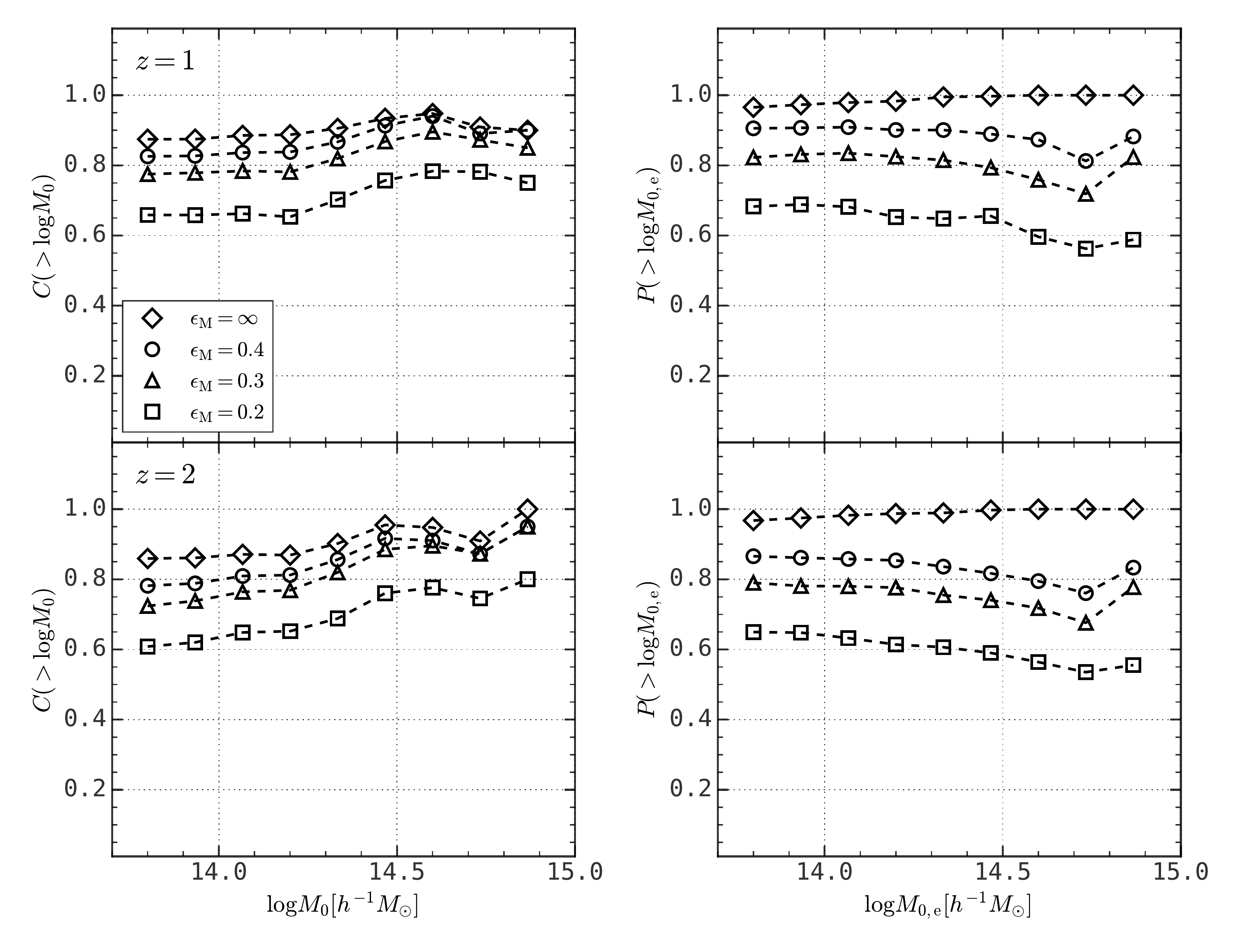}
    \caption{Completeness (\textit{left}) and purity (\textit{right}) of the
        proto-cluster identification at $z=1$ (\textit{upper}) and $z=2$
        (\textit{lower}) as functions of descendant halo mass, $M_0$, with
    different mass error tolerance (see text for detailed explanations). }
    \label{fig:figure/group_level_perf}
\end{figure*}

\begin{figure*}
    \centering \includegraphics[width=0.7\linewidth]{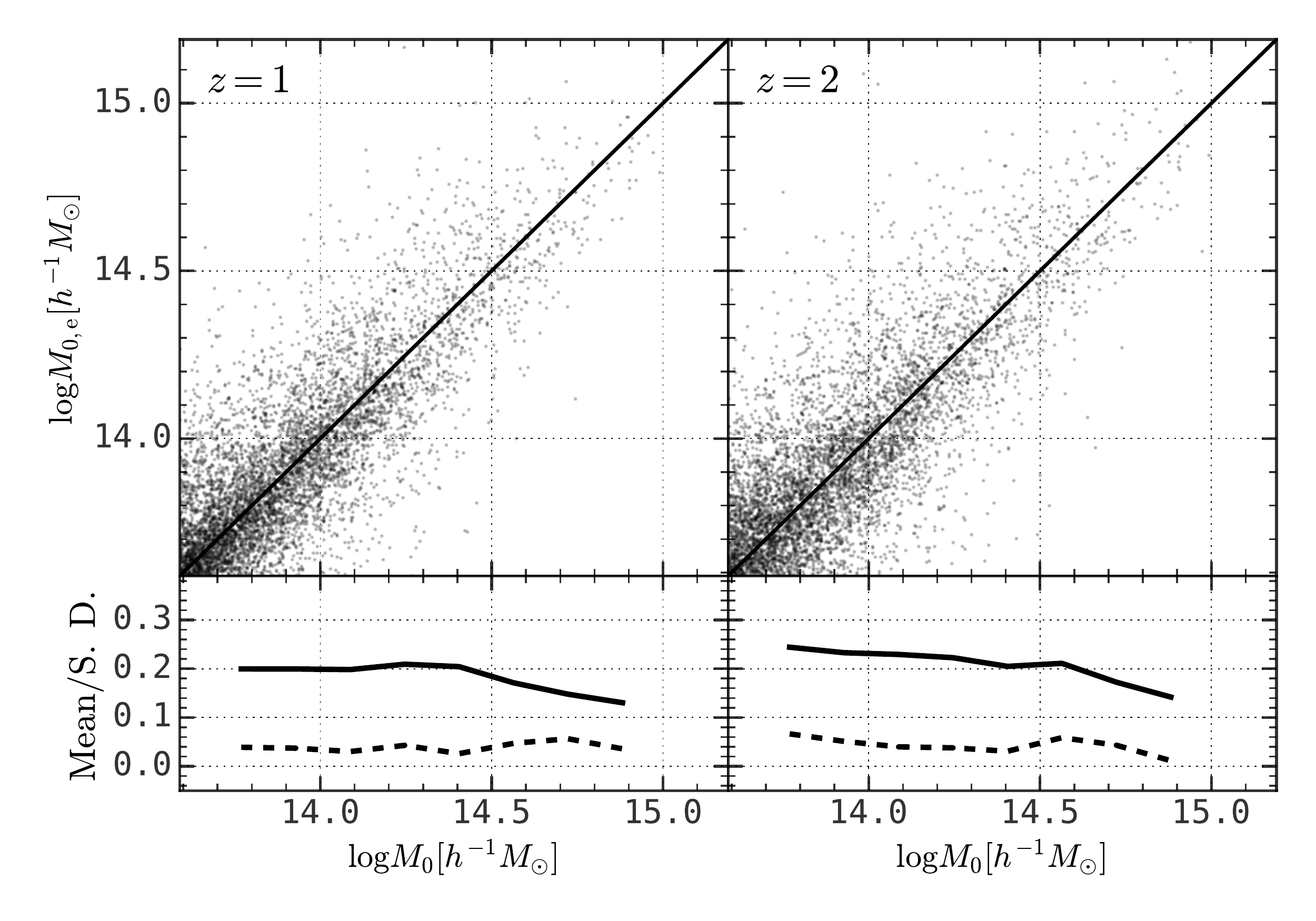}
    \caption{Comparison of the descendant halo mass between the true value and
        the estimated value at $z=1$ (\textit{left}) and $z=2$
        (\textit{right}). The dashed lines in the upper panels are the
        one-to-one line. In the lower panels, the solid and dashed lines show the standard
    deviation and mean of $\log(M_0/M_{0, e})$, respectively.}
    \label{fig:figure/mass_z0_perf}
\end{figure*}

\begin{figure*}
    \centering \includegraphics[width=\linewidth]{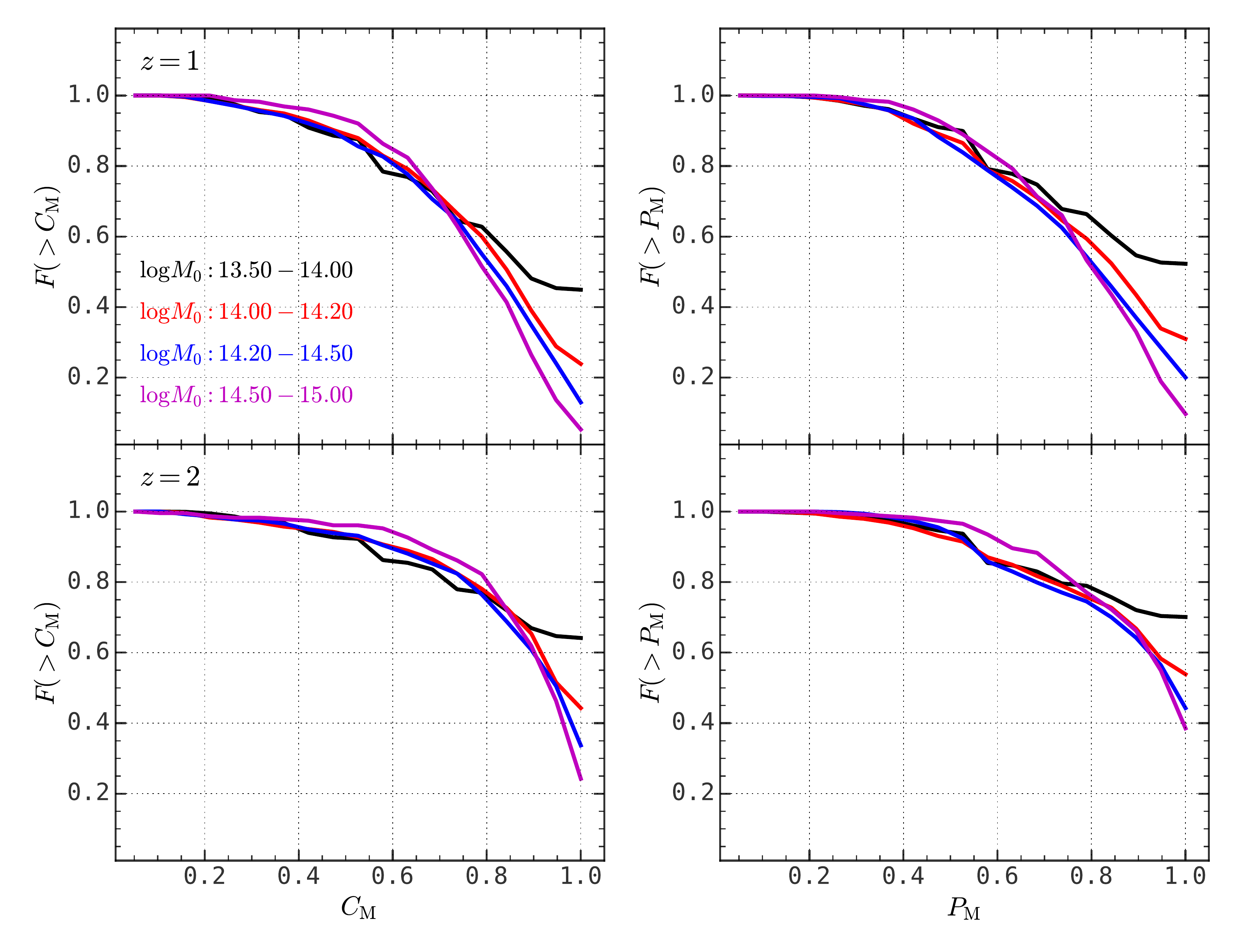}
    \caption{The accumulated membership completeness (\textit{left}) and purity
        (\textit{right}) for the proto-clusters identified and matched with
        true proto-clusters at $z=1$ (\textit{upper panels}) and $z=2$
    (\textit{lower panels}).} \label{fig:figure/member_level_perf}
\end{figure*}

In this subsection, we quantify the performance of our proto-cluster finding
algorithm in terms of the completeness and purity of the identified
proto-cluster population. We define the completeness ($C$) as the fraction of
correctly identified proto-clusters among the true population, and purity ($P$),
as the fraction of the correctly identified proto-clusters among all the
candidate proto-clusters. To do this, we need to define what we mean by a
correct identification of a proto-cluster. We use the most massive halo (MMH)
selected into a candidate proto-cluster to link it to a true proto-cluster.
Thus, if the MMH of the true proto-cluster is the MMH of a candidate
proto-cluster, we then say that the true proto-cluster is correctly identified.
On the other hand, if the MMH of the candidate proto-cluster is the MMH of one
true proto-cluster, the candidate proto-cluster is said to be a correct
identification of a true proto-cluster. In addition, when calculating $C$ and
$P$, we also include a quantity of mass error tolerance, so that the correct
identification of a proto-cluster also requires that
\begin{equation}
    \left|\log\left(M_{0}) - \log(M_{\rm 0, e}\right)\right| <
    \epsilon_M
\end{equation}
where $M_{0}$ is the true descendant halo mass at $z=0$, $M_{0, e}$ is the
estimated descendant halo mass of the matched candidate proto-cluster, and
$\epsilon_M$ is a factor characterizing the mass error tolerance.

In Fig.\,\ref{fig:figure/group_level_perf}, we present the completeness ($C$)
and purity ($P$) for proto-clusters identified at $z=1$ and $2$ as function of
the true descendant halo mass. Results are shown for four different choices of
the mass error tolerance: $\epsilon_M = 0.2, 0.3, 0.4$ and $\infty$, with
$\epsilon_M=\infty$ corresponding to no mass accuracy requirement. We see that,
if no mass accuracy requirement is used, i.e.\ for $\epsilon_M=\infty$, more
than $89\%$ ($87\%$) of the proto-clusters with descendant halo mass above
$10^{14}h^{-1}M_{\odot}$ are identified at $z=1$ ($z=2$), with purity larger
than as $98\%$ ($98\%$). For $\epsilon_M=0.4\,{\rm dex}$, more than $84\%$
(81\%) of the proto-clusters with descendant halo mass above
$10^{14}h^{-1}M_{\odot}$ are identified at $z=1$ ($z=2$), with purity larger
than as $91\%$ ($86\%$). Both $C$ and $P$ decrease by about 20\% when a more
restrictive mass criterion, $\epsilon_M=0.2$, is used.

\subsection{Descendant halo mass}%
\label{ssub:descendant_halo_mass_estimation}

We apply the relations between $M_{\rm tot}$ and $M_0$ obtained in
\S\,\ref{sub:descendant_halo_mass_calibration} to estimate the descendant mass,
$M_{0,e}$ for each identified proto-cluster using the total mass of its member
halos selected by the proto-cluster finder. Fig.\,\ref{fig:figure/mass_z0_perf}
shows the comparison between $M_{0,e}$ and the true descendant halo mass,
$M_0$. As one can see, most of the proto-clusters (black dot) lie close to the
one-to-one line, with the standard deviation (shown by the black solid lines in
lower panels) typically about 0.20 dex and 0.25 dex at $z=1$ and $z=2$,
respectively. In real applications, there are also uncertainties in the masses
assigned to dark matter halos. \citet{wangIdentifyingGalaxyGroups2020} tested
the halo mass estimate using realistic mock catalogs based on the PFS survey
\citep{takadaExtragalacticScienceCosmology2014}, and found that the typical
error in the mass estimate is about 0.2 dex.  Including this uncertainty in the
estimate of $M_{\rm tot}$ increases the standard deviation of $M_{0,e}$ to 0.3
dex (See Fig.\,\ref{fig:figure/mass_z0_perf_compare}).

\subsection{Completeness and purity of member halos}%
\label{ssub:member_level_performance}

Another important performance measurement concerns the member halos identified,
in terms of the fractions of the true member halos and the interlopers, in each
candidate proto-cluster. This kind of performance has been used to test the
galaxy group finder in \cite{yangGalaxyGroupsSDSS2007}. Here we modify the
definition slightly. We define the membership completeness, $C_{\rm M}$, and
membership purity, $P_{\rm M}$, as
\begin{align}
    C_{\rm M} &\equiv N_{\rm ST} / N_{\rm T}\\
    P_{\rm M} &\equiv N_{\rm ST} / N_{\rm S}
\end{align}
where $N_{\rm S}$ is the total number of halos selected into a candidate
proto-cluster, $N_{\rm T}$ is the number of halos in the corresponding true
proto-cluster, and $N_{\rm ST}$ is the number of true member halos in the
candidate proto-cluster. Thus, $C_M=P_M=1$ for a perfect membership
assignments.

The member completeness and purity are shown in Fig.
\ref{fig:figure/member_level_perf}. At $z=1$ ($z=2$), $\sim 60\%$ ($\sim 80\%$)
of the identified proto-clusters have completeness of
$\gtrsim 80\%$, while $\sim 56\%$ ($\sim 80\%$) of them have purity $\gtrsim
80\%$. We note that, in terms of $C_M$ and $P_M$, our proto-cluster finder
performs equally well for different descendant halo mass bins. This suggests
that we can use the same finder for proto-clusters of different descendant halo
masses by re-scaling the selection rule with the descendant halo mass.

\subsection{Conditional mass function of member halos}%
\label{sub:conditional_halo_mass_function}

\begin{figure*}
    \centering \includegraphics[width=0.8\linewidth]{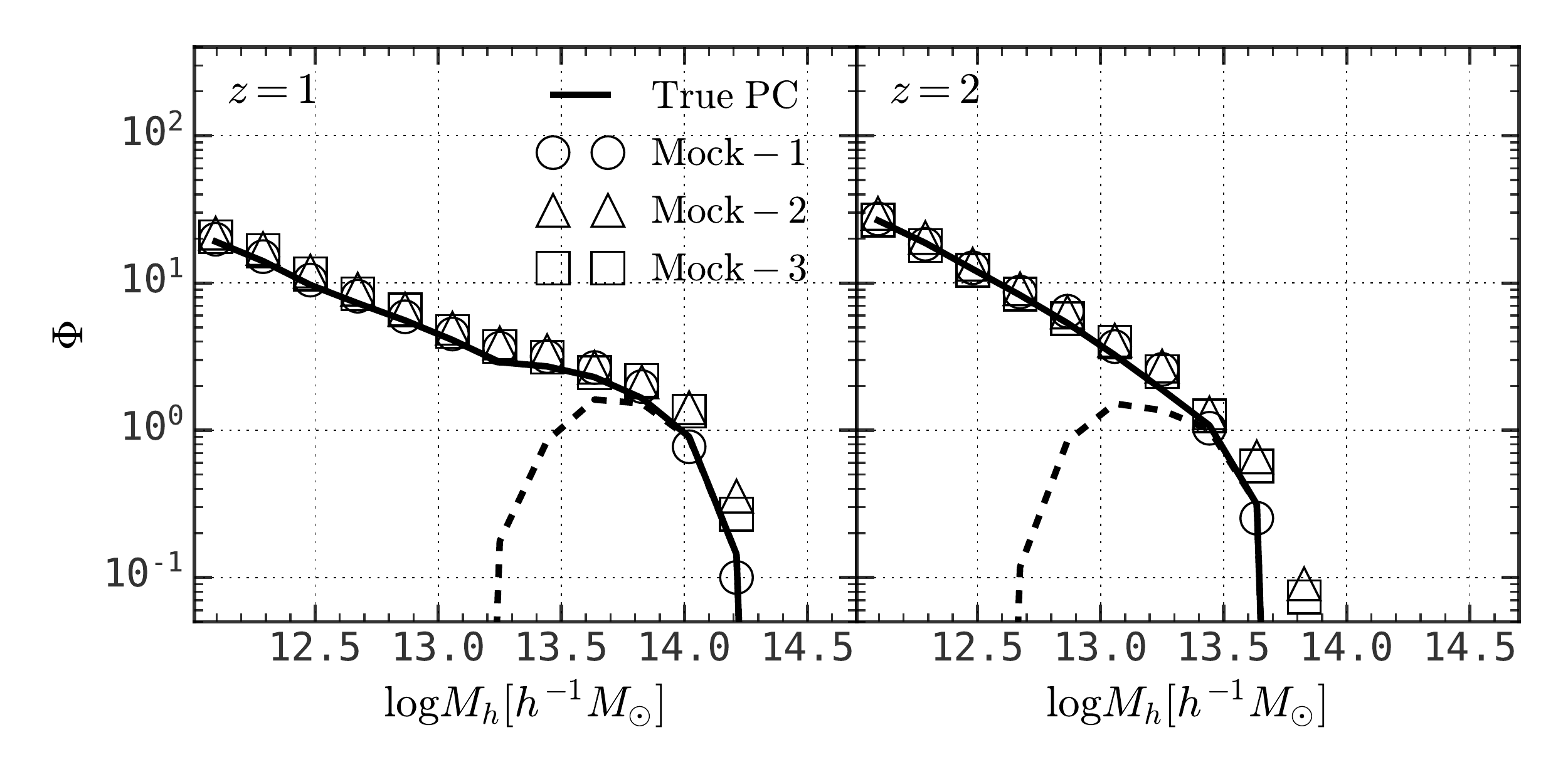}
    \caption{Conditional halo mass function at $z=1$ (\textit{left}) and $z=2$
        (\textit{right}) for proto-clusters with $14.2\leq \log M_{0,
        e}/[h^{-1}M_\odot] \leq 14.5$. The black solid lines are measured from
        the true proto-clusters, the dashed lines are the MMH component of the
        true conditional halo mass function, the circles are measured from the
        candidate proto-cluster catalog. The triangles and squares are results
        for \texttt{Mock-2} and \texttt{Mock-3} defined in
    Appendix~\ref{sec:uncertainties_introduced_by_the_group_finder}.}
    \label{fig:figure/chmf}
\end{figure*}

The conditional halo mass function is defined as
\begin{equation}
    \Phi(M_h|M_{0}^l,M_{0}^u)\equiv \frac{{\rm d}N(M_h |M_{0}\in
    [M_{0}^l,~M_{0}^u])}{N_{\rm PC}~{\rm d}\log M_h}
    \label{chmf}
\end{equation}
where $N(M_h |M_{0}\in [M_{0}^l,~M_{0}^u])$ is the halo mass distribution of
all the halos at redshift $z$ whose descendant halo mass at $z=0$ is in the
range of $[M_{0}^l,~M_{0}^u]$, and $N_{\rm PC}$ is the number of proto-clusters
in that descendant halo mass range. So defined, the conditional halo mass
function describes the average number of halos of a given mass that are
contained in proto-clusters of a given $M_0$. The results of the conditional
halo mass function obtained from our identified proto-clusters are shown as
circles in Fig. \ref{fig:figure/chmf} (Mock-1), and are compared to those
obtained from the true proto-clusters (the solid curves). For comparison, the
dashed curves are for the most massive halos (MMH) in individual
proto-clusters. The conditional halo mass functions above the halo mass limit
of $10^{12}h^{-1}M_{\odot}$ are well reproduced for proto-clusters with $M_0=
10^{14}h^{-1}M_{\odot}$ to $10^{15}h^{-1}M_{\odot}$ at both $z=1$ and $z=2$.
Here we only present the result of one descendant halo mass bin for clarity,
and we can reproduce the function in other bins equally well. Even if we
include a dispersion of 0.20 dex in the estimates of $M_h$, as expected from
the uncertainties produced by the group finder in real applications
\citep{wangIdentifyingGalaxyGroups2020}, the results do not change much (See
appendix.\,\ref{sec:uncertainties_introduced_by_the_group_finder}).

The good match between the recovered conditional halo mass function and the
true one indicates that the member halos of proto-clusters are reliably
identified by our method. Since these halos are the progenitors of sub-halos in
present-day main halos with given $M_0$, and since member galaxies in
present-day galaxy clusters are expected to be connected with the sub-halos,
the proto-clusters and their member halos identified using our method can be
used to link cluster galaxies with their high-$z$ progenitors statistically.

\section{Linking high redshift progenitors to local clusters}%
\label{sec:linking_high_z_progenitors_to_local_clusters}

An important problem in astronomy is to understand how galaxies evolve
with redshift. Various methods have been proposed to link galaxy populations at
different redshifts statistically. For example, one can directly link local
galaxies with their high-$z$ progenitors or link high-$z$ galaxies with
their low-$z$ descendants using a rank-order/abundance matching method
\citep{vandokkumGROWTHMASSIVEGALAXIES2010,behrooziUSINGCUMULATIVENUMBER2013}.
One can also connect galaxies at different redshifts using stellar ages
\citep{cimattiFastEvolvingSize2012}. Finally, one may also study the evolution
of the brightest cluster galaxies by connecting dark matter halos at different
redshifts \citep{lidmanEvidenceSignificantGrowth2012, linSTELLARMASSGROWTH2013,
cookeStellarMassGrowth2019, demaioGrowthBrightestCluster2020}. In this section,
we show how we can use the information provided by proto-clusters to link halos
and galaxies at high-$z$ to their descendants at the present, and vise versa.

\subsection{High redshift dark matter halos and their descendants}%
\label{sub:linking_halos}

\begin{figure*}
    \centering
    \includegraphics[width=1\linewidth]{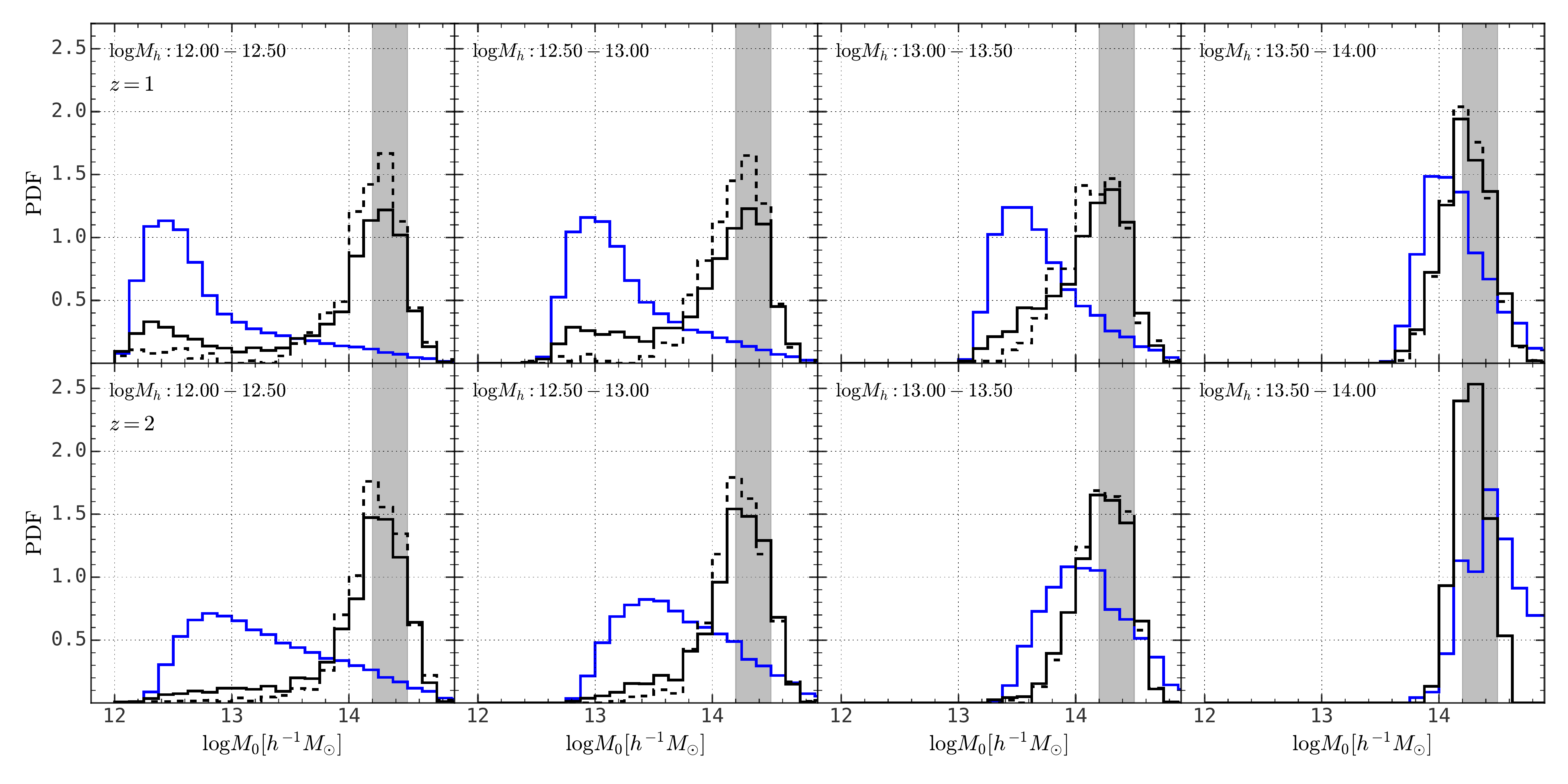}
    \caption{True descendant halo mass distribution for selected dark matter
        halos at high $z$. Four \textit{columns} are for dark matter halos
        selected in different mass bins: $\log(M_h/[h^{-1}M_\odot]) \in$
        [12.0, 12.5],~[12.5, 13.0],~[13.0, 13.5] and [13.5, 14.0]. The
        \textit{blue} histogram shows the true descendant halo mass
        distribution in each halo mass bin. The \textit{black solid} histogram
        shows the true descendant halo mass distribution with the additional
        requirement $14.2 \leq\log(M_{0, e} /[h^{-1}M_\odot])\leq 14.5$. The
        \textit{black dashed} histogram shows the true descendant halo mass
        distribution after eliminating halos in the outskirts of the candidate
        proto-clusters. The gray
        region indicates the range of $M_{0, e}$. The \textit{upper} panels
        are for $z=1$ and the \textit{lower} panels are for $z=2$.
    }%
    \label{fig:figure/hm_z0_halos}
\end{figure*}

Let us first examine how halos selected at high $z$ are linked to halos at
$z=0$. To do this, we select all halos of a given mass at a given high $z$ and
study the mass distribution of their descendants at $z=0$. The blue histograms
in Fig.\,\ref{fig:figure/hm_z0_halos} show the descendant mass ($M_0$)
distribution for halos identified at $z=1$ and $z=2$, respectively. For a given
halo mass bin, $M_1\le M_h\le M_2$, the $M_0$ distribution is peaked at a value
that is a couple of times larger than the average mass of the halos selected.
Most of the low-mass halos will end up in relatively low-mass descendant halos
at $z=0$. However, there is an extended tail towards high $M_0$, which is
expected to be dominated by halos in massive proto-clusters. We can use the
information provided by the identified proto-clusters to refine the connection
between high-$z$ halos and their $z=0$ descendants. To this end, we split halos
into several bins of the estimated descendant mass ($M_{0,e}$), and obtain the
true descendant mass ($M_0$) distributions for halos in each $M_{0,e}$ bin. As
an example, the black histograms in the figure show these
distributions for $14.2\leq \log M_{0, e}/[h^{-1}M_\odot]\leq 14.5$ (indicated
by the vertical gray bands). Results for other bins are qualitatively the same
and not shown here. One can see that the peak of the distribution is now
roughly at the gray region, indicating that the use of the constraint on
$M_{0,e}$ can effectively select the progenitors of the halos at $z=0$. There is,
however, a long tail at the low descendant mass
end, which is contributed by interloper halos in the identified proto-clusters.
\footnote{An interloper is a halo that does not belong to the
proto-cluster which the halo is assigned to. It may be a member of another
proto-cluster if the two proto-clusters are close to each other.} We can
reduce the contribution of these interlopers by eliminating halos in the
outskirts of the identified proto-clusters. To do this, we first define two
effective radii,
\begin{align}
    R_{e,\parallel} &= \sqrt{\sum_i M_{h, i}(X_i - {\bar X})^2 \over M_{\rm tot}} \\
    R_{e,\perp} &= \sqrt{\sum_i M_{h, i}\left[(Y_i - {\bar Y})^2 + (Z_i - {\bar Z})^2\right] \over M_{\rm tot}} \\
    {\bar \mu} &= {\sum_i M_{h, i} {\mu_i} \over M_{\rm tot}}, ~~\mu=X,~Y,~Z
\end{align}
for each identified proto-cluster, where $M_{h,i}$ is the mass and $(X_i,~Y_i,~
Z_i)$ is the location of the $i$-th halo in the candidate proto-cluster, $M_{\rm
tot}=\sum_i M_{h,i}$ is the sum of the halo mass in the proto-cluster, and
$({\bar X},~{\bar Y},~{\bar Z})$ is the mass-weighted center. Note that we have
assumed that $X$ is in the line-of-sight direction. We only keep member halos
that satisfy
\begin{align}
    |X_i - \bar{X}| &< R_{e, \parallel} \\
    \sqrt{(Y_i - \bar{Y})^2 + (Z_i - \bar{Z})^2} &< R_{e, \perp}.
\end{align}
The black dashed histograms in Fig.\,\ref{fig:figure/hm_z0_halos} show the
corresponding distributions. As one can see, the long tail in the low mass end
of the distribution is significantly suppressed. All these demonstrate that the use
of the proto-clusters identified can effectively improve the link between
high-$z$ halos and their descendants at $z=0$.

\subsection{Linking high redshift galaxies to their descendants}%
\label{sub:linking_galaxies}

Proto-clusters also provide us a statistical link between galaxies at
high-$z$ and those in the local Universe, thereby allowing us to study the
time evolution of the galaxy population. Here we use all galaxies, produced
by the empirical model (see \S\,\ref{sub:the_simulation_and_empirical_model}),
with $M_*\geq 10^{10}h^{-1}M_\odot$ in halos with $M_h\geq
10^{12}h^{-1}M_\odot$ at $z=1$ and $z=2$, together with their $z=0$ descendants
identified by following their merging trees.

\subsubsection{The conditional stellar mass function of galaxies in proto-clusters}%
\label{ssub:the_conditional_stellar_mass_function_of_galaxies_in_proto_clusters}

\begin{figure*}
    \centering \includegraphics[width=1\linewidth]{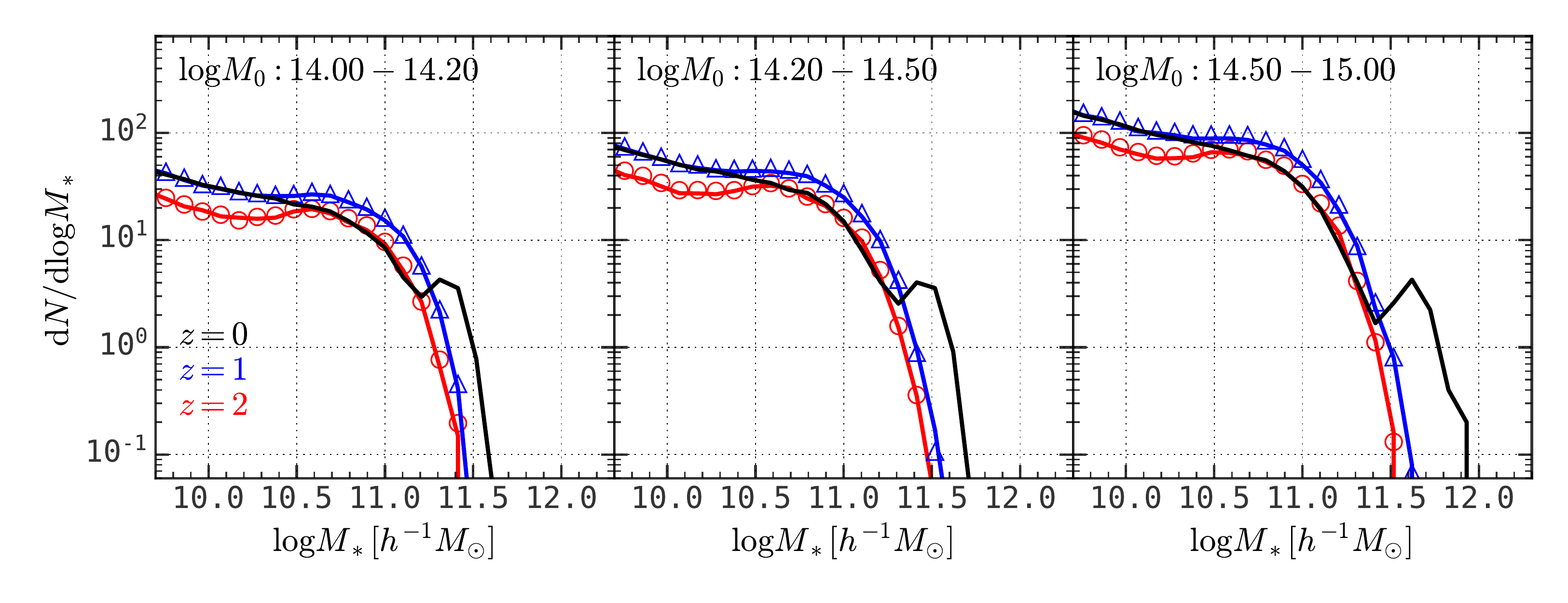}
    \caption{Proto-cluster conditional stellar mass function. The three panels
        are for different descendant halo mass bins. In each panel, the colored
        solid lines are results for true proto-clusters and the symbols are for
        proto-clusters identified with our method. Results are shown for $z=1$
        (\textit{blue}) and $z=2$ (\textit{red}). For comparison, the black
        line shows the conditional stellar mass function of galaxies in the
        $z=0$ descendant halos.
    }%
    \label{fig:figure/csmf}
\end{figure*}

First we examine member galaxies of proto-clusters by studying the conditional
stellar mass function of proto-cluster galaxies (hereafter CSMF of PCG), and
compare it to the CSMF of cluster galaxies (hereafter CSMF of CG) at $z=0$. The
CSMF of PCG is defined as
\begin{equation}
    \Phi_*(M_*|M_{0}^l, M_{0}^u)\equiv \frac{{\rm d}N(M_* |M_{0}\in
    [M_{0}^l,~M_{0}^u])}{N_{\rm PC}~{\rm d}\log M_*}
\end{equation}
where $N(M_* |M_{0}\in [M_{0}^l,~M_{0}^u])$ is the stellar mass distribution of
all the galaxies at high redshift contained in proto-clusters with $M_0^l\le
M_0\le M_0^u$, and the normalization factor, $N_{\rm PC}$, is the number of
proto-clusters in the same $M_0$ range. Fig.\,\ref{fig:figure/csmf} shows the
CSMF of PCG measured from the true proto-clusters as solid lines, with blue for
$z=1$ and red for $z=2$. The result obtained from the identified
proto-clusters, shown in circles, matches that for the true proto-clusters very
well, indicating that the identified proto-clusters can be used to represent
the galaxy population in the true proto-clusters reliably. The black curves are
the CSMFs of CG at $z=0$. As one can see, the number of galaxies increases
between $z=2$ and $z=1$ by a factor of about two over the entire stellar mass
range. Between $z=1$ and $z=0$, the number of galaxies at
$M_* < 10^{10.5} h^{-1} M_\odot$ changes little, the number around $M_* =
10^{11} h^{-1} M_\odot$ decreases by a factor of $\sim 1.5$, while the number
at the massive end increases by a large amount. Note that the peak at the
massive end is dominated by central galaxies in $z=0$ clusters. Such a feature
is absent at higher $z$, indicating that the build-up of the mass of central
galaxies by accretion happens mostly below $z=1$. Clearly, such conditional
stellar mass functions carry important information about the evolution of the
galaxy population in clusters of galaxies.

\subsubsection{Descendant mass distribution}%
\label{ssub:descendant_mass_distribution}

\begin{figure*}
    \centering
    \includegraphics[width=1\linewidth]{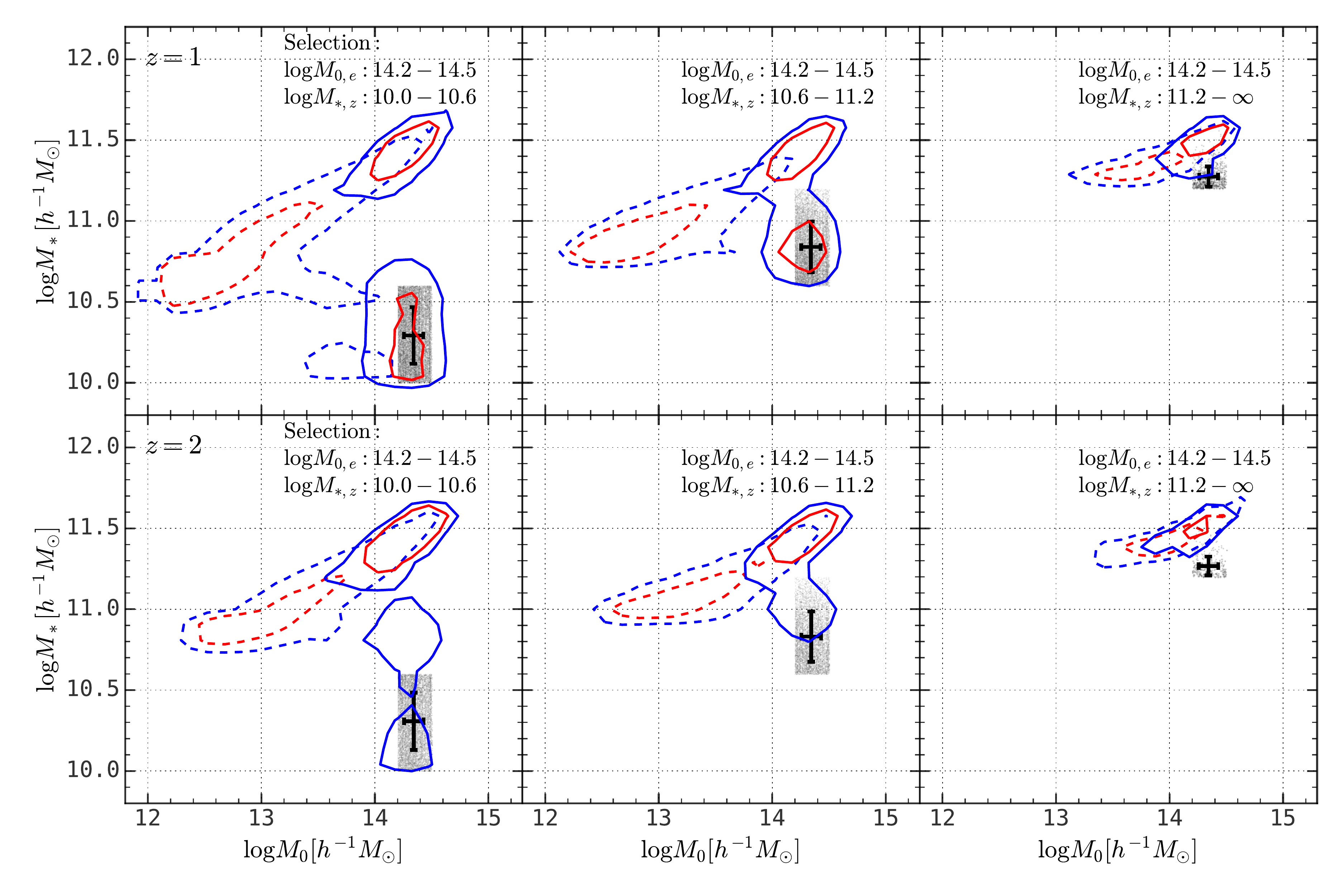}
    \caption{
        Joint distribution in descendant halo mass and descendant stellar mass.
        The three \textit{columns} are for galaxies in different stellar mass
        bins: [10.0, 10.6], [10.6, 11.2], and [11.2, $\infty$] selected at high
        $z$. The \textit{dashed} and \textit{solid} contours both show the
        joint distribution in the true descendant halo mass ($M_0$) and
        descendant stellar mass ($M_{*, 0}$) distribution, with the dashed ones
        for galaxies selected only with stellar mass ($M_*$), and the solid
        ones for galaxies selected with
        both the stellar mass ($M_*$) and the additional requirement that
        $14.2\leq \log (M_{0, e}/[h^{-1}M\odot])\leq 14.5$, indicated by the
        black points. The
        error bar shows the mean and standard deviation of the black points.
        The \textit{blue} and \textit{red} contours enclose 80\% and 50\% of
        the galaxies, respectively. The \textit{upper} panels are for $z=1$ and
        the \textit{lower} panels are for $z=2$.
    }%
    \label{fig:figure/desc_gal_prop_combined}
\end{figure*}

\begin{figure*}
    \centering
    \includegraphics[width=1\linewidth]{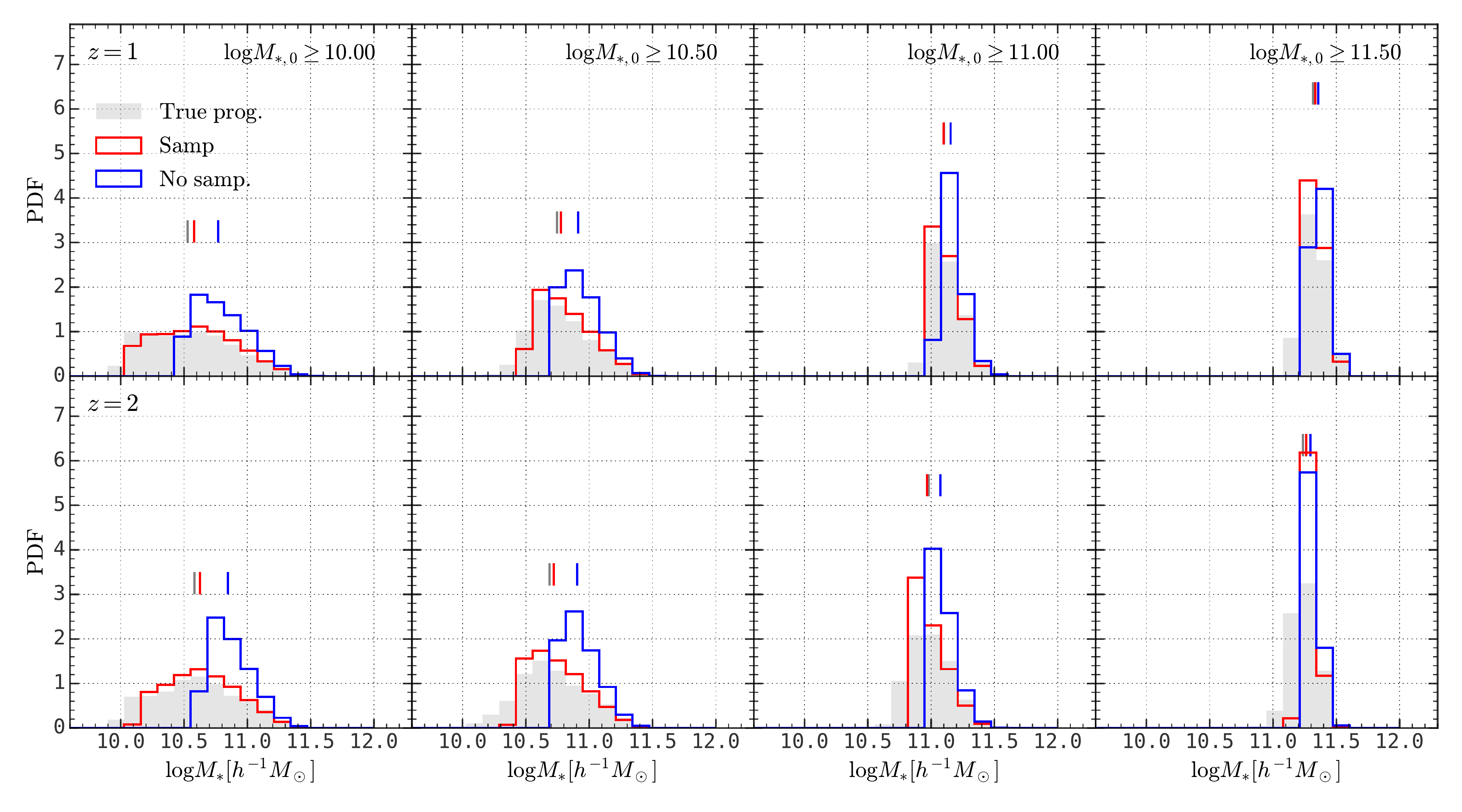}
    \caption{Stellar mass distribution of the most massive progenitor galaxies
        at $z=1$ (\textit{upper}) and $z=2$ (\textit{lower}) for galaxies
        selected at $z=0$ with different stellar mass cuts (\textit{different
        columns}). The \textit{gray} histograms show the true distribution
        extracted from the galaxy merger tree, while the \textit{red} and
        \textit{blue} are the results using rank matching method with or
        without the sampling process, respectively (See text). The short
        colored bars in each panel indicate the medians of the histograms of
        the corresponding colors.
    }%
    \label{fig:figure/galaxy_prog_hist_with_am}
\end{figure*}

\begin{figure*}
    \centering
    \includegraphics[width=0.8\linewidth]{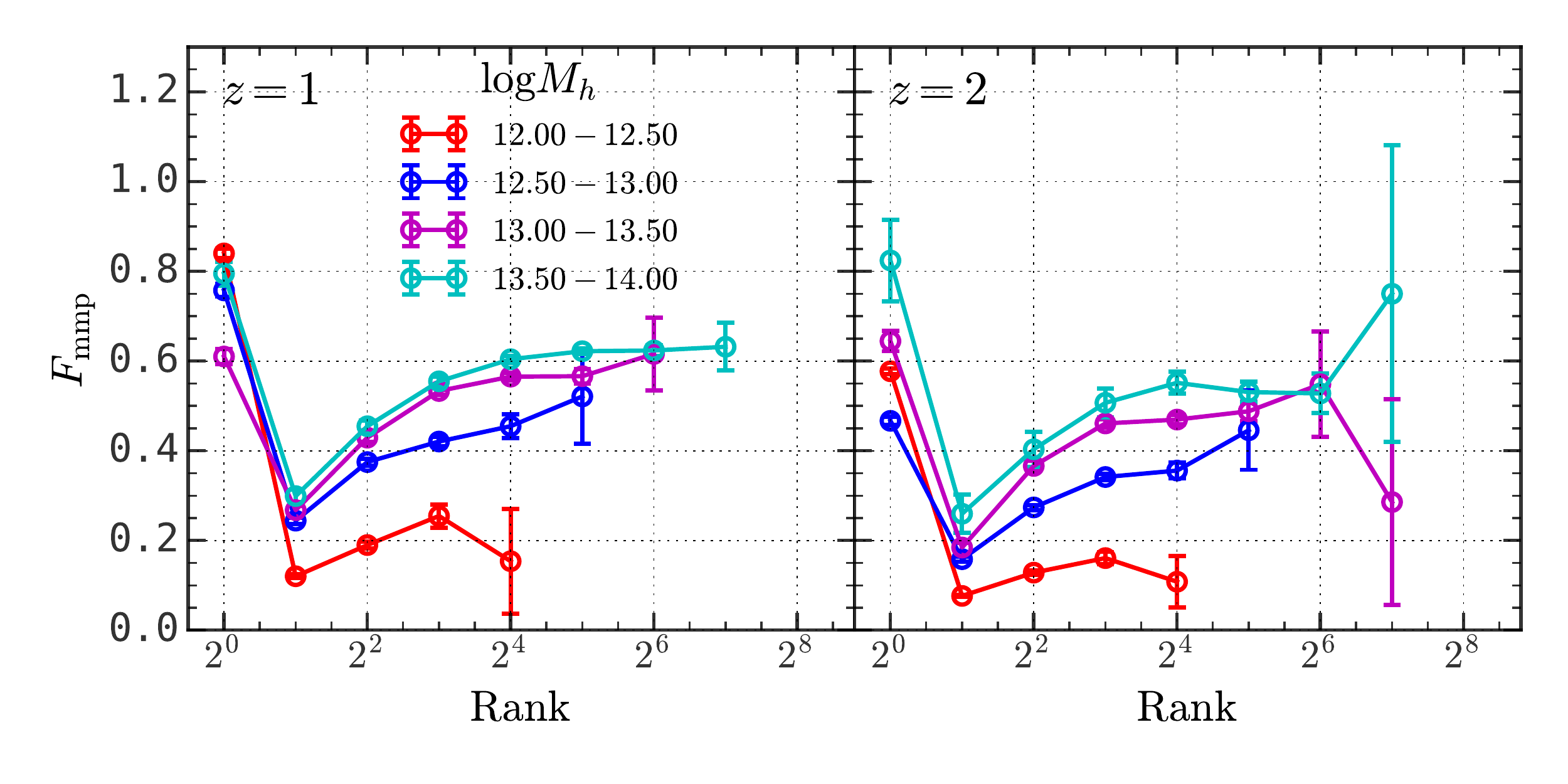}
    \caption{Fraction of the most massive progenitors, $F_{\rm mmp}$, as a function
        of stellar mass rank in their host halo for galaxies at $z=1$
        (\textit{left}) and $z=2$ (\textit{right}). Different colors are
        for different halo mass bins.
    }%
    \label{fig:figure/mmp_probability}
\end{figure*}

We first examine the descendants of galaxies selected at high $z$. To this end,
We select all galaxies at a given $z$ in stellar mass bins and examine their
descendant galaxies at $z=0$. In Fig.\,\ref{fig:figure/desc_gal_prop_combined},
the dashed contours show the distribution of the descendants in the stellar mass
($M_{*,0}$) versus halo mass ($M_0$) space. Results are shown for galaxies
selected at $z=1$ and $z=2$ in three bins of $M_*$:
$\log(M_*/[h^{-1}M_\odot])\in$ [10.0, 10.6], [10.6, 11.2], [11.2, $\infty$],
corresponding to the three columns of the figure. As one can see, if
restriction is only imposed on the stellar mass of high-$z$ galaxies, the
descendants mostly reside in low-mass halos, particularly for the low $M_{*}$
bins.

Since galaxy properties in the local universe are observed to depend strongly
on their host halo mass \citep[e.g.][]{weinmannPropertiesGalaxyGroups2006,
wangELUCIDIVGalaxy2018}, it is interesting to focus on galaxies that end up in
halos/clusters of a given mass in the local Universe. We thus present the joint
distribution in $(M_{*, 0},~M_0)$ for galaxies of given $M_*$ that are
contained in proto-clusters of given estimated descendant halo mass, $M_{0,
e}$. In Fig.\,\ref{fig:figure/desc_gal_prop_combined}, the black points show
the distribution of $(M_*,~M_{0, e})$ for the high-$z$ galaxies, while the solid
contours show the descendant distributions in $(M_{*, 0},~M_0)$. Results are
only shown for $14.2 \leq
\log\left(M_{0,e}/\left[h^{-1}M_\odot\right]\right)\leq 14.5$; results for
other $M_{0,e}$ bins are similar and omitted for brevity. Here one can see that
the descendant halo mass distribution matches the input range of $M_{0, e}$.
This is expected, because the proto-clusters identified with our method
correspond to their descendants accurately.

It is interesting to note that, for low-$M_*$ galaxies at $z=1$, the
distribution of their descendant galaxies shows two peaks and a middle valley,
while for $z=2$, the distribution shows three peaks. We believe that these
three populations correspond to three evolution tracks. Galaxies in the low
mass peak have been quenched and experienced no major mass acquisitions since
$z=1$ or $z=2$, so that their stellar mass does not grow much. The middle
valley at $z=1$ or the middle peak at $z=2$ corresponds to central galaxies of
low mass halos, in which the stellar mass grows steadily through star formation
before they are accreted into the cluster at later time. Finally, the high
mass peaks correspond to high-$z$ galaxies that have merged with massive
galaxies by $z=0$, so that their descendant stellar mass is much larger
than their stellar mass at high $z$. The low-mass component is more prominent
at $z=1$ compared with $z=2$, because the time interval available for star
formation and merger is shorter. For the most massive galaxies, the increase of
the stellar mass in the descendants is modest, about a factor of 1.5 to 2. We
note that the results here are based on the empirical model
\citep{luEmpiricalModelStar2014, luStarFormationStellar2015,
chenELUCIDVICosmic2019}. However, the evolution tracks of cluster galaxies
discussed above are expected to be valid in the general paradigm of galaxy
formation.

\subsubsection{Abundance matching}%
\label{ssub:abundance_matching}

Next, we examine the progenitors of the galaxies in $z=0$ clusters. Because of
merger, each galaxy at $z=0$ may correspond to more than one galaxy at high $z$.
Thus, we consider only the most massive progenitors.
Fig.\,\ref{fig:figure/galaxy_prog_hist_with_am} shows the stellar mass
distribution of the most massive progenitors at $z=1$ (upper panels) and $z=2$
(lower panels) for $z=0$ galaxies with different stellar masses, $M_{*,0}$. For
illustration, results are shown for halos/clusters with $M_{0}\geq 10^{14}h^{-1}
M_\odot$. The gray filled histogram shows the distribution of true progenitor
galaxies identified with the galaxy merging trees in the empirical model.
Our goal is to recover this distribution from the proto-clusters identified
with our method by linking cluster galaxies with the most massive progenitors
at high $z$.

As a first attempt, we use a simple abundance matching scheme, assuming that
more massive galaxies at $z=0$ have more massive progenitors at high $z$
\citep{vandokkumGROWTHMASSIVEGALAXIES2010, behrooziUSINGCUMULATIVENUMBER2013}.
We first separate galaxies at a given high $z$ according to the estimated
descendant halo mass, $M_{0, e}$. For each galaxy at $z=0$, we then match it
with a high-$z$ galaxy that has the same stellar mass rank in the same $M_{0,
e}$ bin. The blue histograms in Fig.\,\ref{fig:figure/galaxy_prog_hist_with_am}
show the results of the progenitor stellar mass distribution obtained from this
scheme. For high $M_{*,0}$, the scheme reproduces the distribution quite well.
For low $M_{*,0}$, however, the distribution is biased towards the massive end
relative to the true distribution. This bias is caused by galaxy merging. As
shown in Fig.\,\ref{fig:figure/desc_gal_prop_combined}, many high-$z$ galaxies
with intermediate stellar masses have merged into massive descendants by $z=0$.
A fraction of these galaxies should not be used in the abundance matching, as
they are not the most massive progenitors of any galaxies at $z=0$.

To deal with the problem caused by galaxy merging in the abundance matching, we
need to exclude, in the abundance matching scheme, galaxies that are not the
most massive progenitors of any galaxies at $z=0$. In real applications, this
can be done only in a statistical sense, as one cannot establish the merger
trees for individual galaxies in observation. Statistically, we can estimate
the fraction of the most massive progenitors among all progenitors of given
properties: $F_{\rm mmp}(z) =N_{\rm mmp}(z)/N_{\rm all}(z)$, where $N_{\rm
mmp}(z)$ is the number of the most massive progenitors at $z$ for galaxies at
$z=0$, and $N_{\rm all}(z)$ is the number of all galaxies. This fraction can be
estimated from our empirical model or from numerical simulations, and the hope
is that it can be presented in a way such that it does not depend on galaxy
formation model strongly (see
Appendix~\ref{sec:abundance_matching_for_illustris_tng} for a test). To
achieve this, we first divide galaxies at a given high $z$ into bins of their
host halo masses, $M_h$. Each galaxy is assigned a rank according to its
stellar mass rank in its halo, with the first rank corresponding to the most
massive galaxy and so on. $F_{\rm mmp}$ is estimated in each $(M_h,~{\rm
rank})$ bin and presented in Fig.\,\ref{fig:figure/mmp_probability} for $z=1$
(left panel) and $z=2$ (right panel). Once $F_{\rm mmp}$ is known, we can
randomly select galaxies at high $z$ as the most massive progenitor of a galaxy
at $z=0$ with a probability $F_{\rm mmp}$. The abundance matching scheme can
then be used between $z=0$ galaxies and the galaxies in the random sample of
the most massive progenitors to establish, statistically, connections between
$z=0$ cluster galaxies and their progenitors. The red histograms in
Fig.\,\ref{fig:figure/galaxy_prog_hist_with_am} show the mass distributions of
the most massive progenitors matched in this way. These distributions match the
true distribution (the shaded histograms) well, indicating that our method
provides a statistically reliable way to link galaxies to their progenitors. As
shown in Appendix~\ref{sec:local_brightest_central_galaxies_and_their_progenitor_galaxies},
our method also performs better than those used in the literature to link the
brightest central galaxies (BCGs) to their progenitors. Note that different
realizations of $F_{\rm mmp}$ lead to different samples of the most massive
progenitors. The variance among these samples provides a useful measure of the
uncertainty in the abundance matching scheme. Finally, we note that the
distribution of $F_{\rm mmp}$ shown in Fig.\,\ref{fig:figure/mmp_probability}
is bimodal. Galaxies of $\rm Rank=1$ are the most massive centrals in their
halos, and are expected to experience different mergers than satellites in the
subsequent evolution.

\section{Summary}%
\label{sec:summary}

In this paper, we develop a method to identify proto-clusters from halos/groups
identified in galaxy surveys at high $z$. We demonstrate how the information
provided by groups and proto-clusters can be used to establish the connections
of cluster galaxies in the present-day universe to their high-$z$ progenitors.
Our proto-cluster finder is based on an extension of the traditional FoF
algorithm applied to dark matter halos represented by galaxy groups/clusters.
Compared with previous methods of proto-cluster identification, our method does
not depend on details of how galaxies form in dark matter halos. Our main
results can be summarized as follows.

\begin{enumerate}
     \item

        Using samples of halos and galaxies in simulations, we find that our
        proto-cluster finder can identify $\gtrsim 85\%$ of the true
        proto-clusters with purity $\gtrsim 95\%$. The standard deviation in
        the descendant halo mass estimate is smaller than $0.25$ dex.

    \item

        For the assignments of member halos to proto-clusters, our test
        shows that $\sim 70\%$ of the candidate proto-clusters have both
        completeness and purity $\gtrsim 80\%$ in halo memberships.

    \item

        We show that the proto-clusters identified by our method provide
        important information to link halos and galaxies across different
        redshifts. With the help of proto-clusters, one can effectively select
        halos and galaxies at high $z$ that are progenitors of clusters and
        cluster galaxies at the present day. This can help us understand the
        evolution history for galaxies in local clusters.

    \item

        We find that the mass function of member halos and the stellar mass
        function of member galaxies in true proto-clusters are well reproduced
        by the proto-clusters selected with our method.

    \item

        The comparison of the galaxy population in proto-clusters
        with that in present-day clusters carries important information about
        the evolution of cluster galaxies. We find that relatively low-mass
        galaxies in proto-clusters in general can be divided into three
        different populations: galaxies whose stellar mass changed little
        during the redshift range in question; galaxies that have increased
        their stellar mass significantly by star formation before quenched by
        the cluster environment; galaxies that have merged into more massive
        galaxies. Massive galaxies typically increase their stellar mass by
        accreting lower-mass galaxies.

    \item

        We develop an abundance matching method to connect galaxies in
        proto-clusters with their descendants in present-day clusters, taking into account
        the bias produced by mergers of galaxies. We find that the probability
        for a high-$z$ galaxy in a proto-cluster to be the most massive
        progenitor of a cluster galaxy at the present day can be calibrated
        reliably in a way without depending on the details of the galaxy
        formation process. Our test shows that this probability can be used to
        successfully recover the progenitor stellar mass distribution for
        galaxies in local clusters.

\end{enumerate}

Our method can be applied straightforwardly to real surveys of high-$z$
galaxies, such as zCOSMOS \citep{lillyZCOSMOS10kBRIGHTSPECTROSCOPIC2009}, PFS
\citep{takadaExtragalacticScienceCosmology2014}, and any other surveys from
which galaxy groups/clusters can be identified reliably to represent the dark
matter halo population. We have tested the impact of a number of general
uncertainties, such as redshift-space distortion, incompleteness of
groups/clusters, and uncertainties in halo mass estimates, and found that our
method works reliably under the influences of these uncertainties. In real
applications, we may still need to test the method using realistic mock
catalogs to quantify the impact of selection effects in a specific survey, but
this is straightforward to do. With the advent of large surveys of high-$z$
galaxies, we expect that our method will provide a new avenue to investigate
the formation and evolution of clusters of galaxies as well as the evolution of
their galaxy populations.

\section*{Acknowledgements}

This work is supported by the National Key R\&D Program of China (grant No.
2018YFA0404502, 2018YFA0404503), and the National Science Foundation of China
(grant Nos. 11821303, 11973030, 11673015, 11733004, 11761131004, 11761141012).
We acknowledge Dandan Xu, Yuning Zhang and Jingjing Shi for accessing the TNG
simulation data. Kai Wang and Yangyao Chen gratefully acknowledge the financial
support from China Scholarship Council.

\section*{Data availability}

The data products of this article will be available on requests to the
corresponding author. The computation was supported by the HPC
toolkit \specialname[hipp] at \url{https://github.com/ChenYangyao/hipp}.

\bibliographystyle{mnras}
\bibliography{bibtex.bib}


\appendix

\section{Uncertainties introduced by the group finder}%
\label{sec:uncertainties_introduced_by_the_group_finder}

\begin{figure*}
    \centering
    \includegraphics[width=0.9\linewidth]{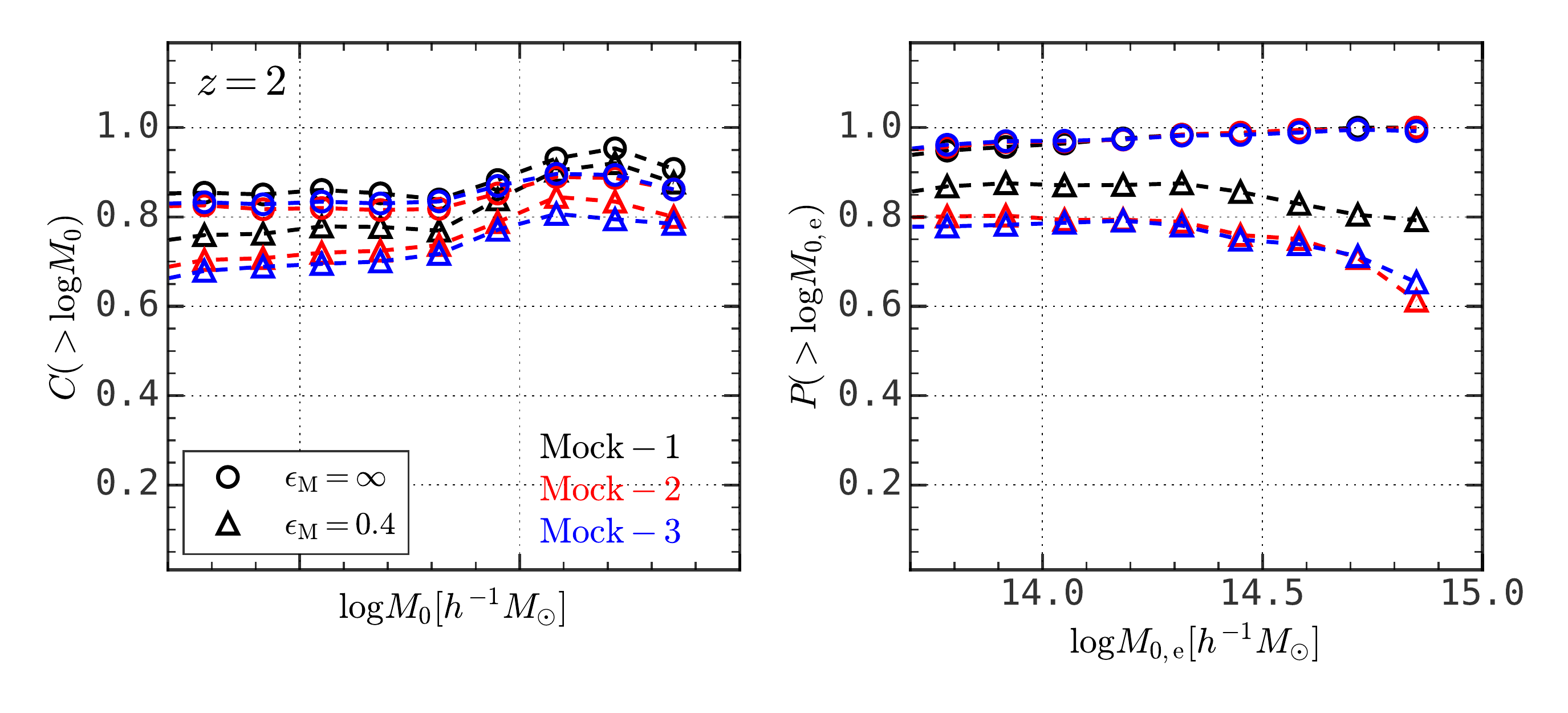}
    \caption{
        Same as Fig.\,\ref{fig:figure/group_level_perf} at $z=2$. The black
        lines are for \texttt{Mock-1}, the red lines are for \texttt{Mock-2}
        and the blue lines are for \texttt{Mock-3}.
    }%
    \label{fig:figure/group_level_performance_compare_z2}
\end{figure*}

\begin{figure*}
    \centering
    \includegraphics[width=0.8\linewidth]{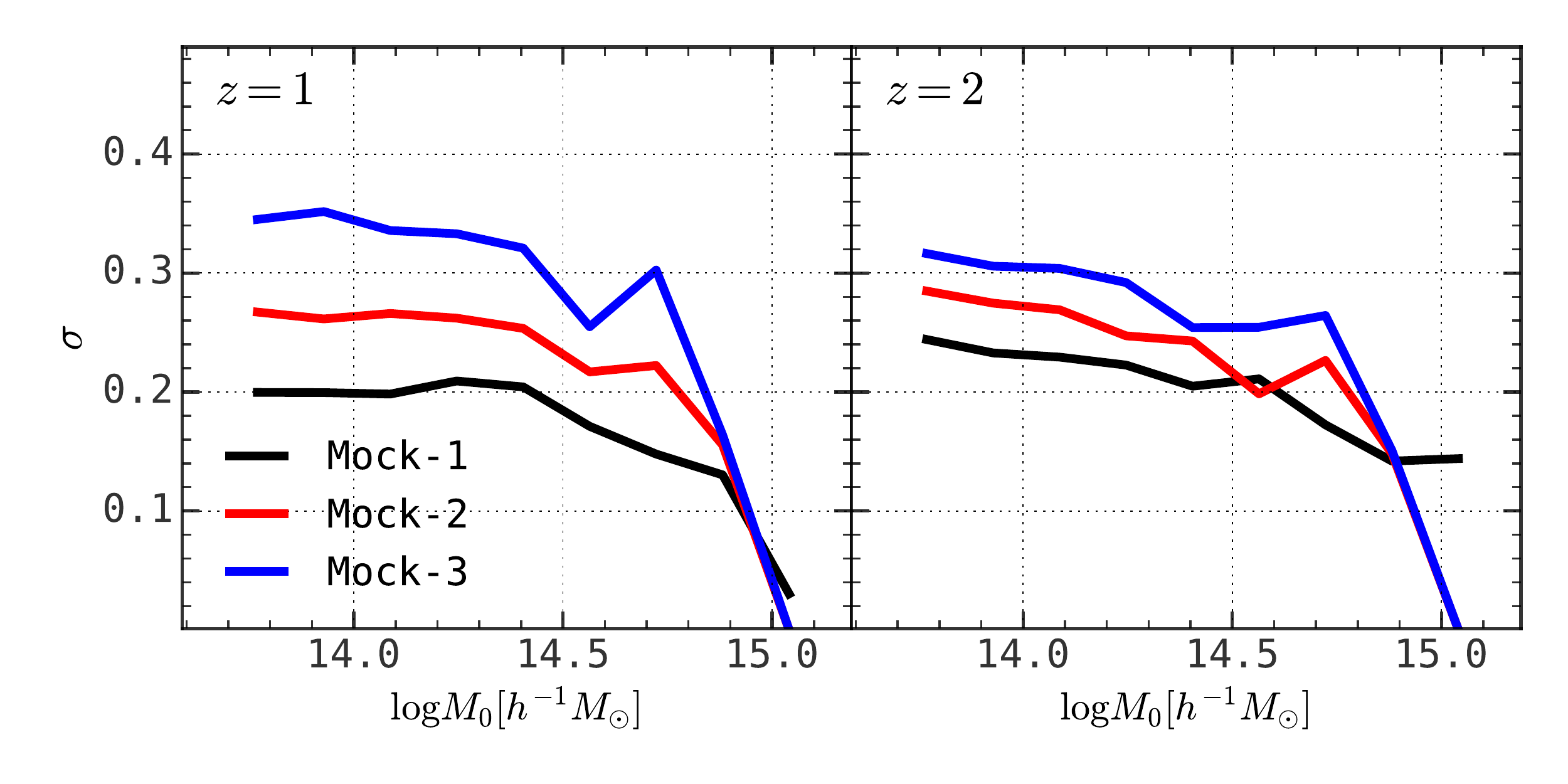}
    \caption{Same as the lower panels in Fig.\,\ref{fig:figure/mass_z0_perf}.
        The black lines are for \texttt{Mock-1}, the red lines are for
        \texttt{Mock-2} and the blue lines are for \texttt{Mock-3}.
    }%
    \label{fig:figure/mass_z0_perf_compare}
\end{figure*}

As shown in \citet{wangIdentifyingGalaxyGroups2020}, a well-designed group
finder applied to high-$z$ redshift surveys such as the PFS can achieve a
completeness of $\gtrsim 90\%$ with a halo mass uncertainty of about $0.2$ dex
for galaxy groups/clusters above $10^{12}h^{-1}M_{\odot}$. These uncertainties
will also affect the performance of the proto-cluster identification, as our
method is based on halos. Here we employ two additional mocks to mimic
uncertainties introduced by the galaxy group finding process. Thus we make
comparisons between the following three mocks:
\begin{itemize}
    \item \texttt{Mock-1}: The same as the mock used in the main part of the
        paper.
    \item \texttt{Mock-2}: Add the following uncertainty to halo mass:
        \begin{equation}
            \log M_h \leftarrow \log M_h + N(0,~0.2)
        \end{equation}
        where $N(0,~0.2)$ is a random number generated from a Gaussian
        distribution with mean $\mu=0$ and dispersion $\sigma=0.2$.
    \item \texttt{Mock-3}: Add the same halo mass uncertainty as in
        \texttt{Mock-2}. In addition we randomly drop $10\%$ of the halos to
        mimic the incompleteness produced by the group finding process.
\end{itemize}
Fig.\,\ref{fig:figure/group_level_performance_compare_z2} shows the result of
group-level performance at $z=2$. We can see that the completeness and purity
for $\epsilon_M=\infty$ are nearly unchanged, while the performance decreases
for $\epsilon_M=0.4$, which is caused by the error in the descendant halo mass
calibration. Here we only show the result of $z=2$ for brevity since the result
of $z=1$ has a similar trend. For the member level performance, the results is nearly
the same as those shown in Fig.\,\ref{fig:figure/member_level_perf}, so we also omit it. This
result also verifies that the decreasing of the group-level performance is caused
by the descendant halo mass calibration, instead of the member halo assignment. We also
present the standard deviation for the descendant halo mass calibration in
Fig.\,\ref{fig:figure/mass_z0_perf_compare}. One can see that the error is
increased by both the halo mass uncertainty and halo incompleteness. The
conditional halo mass functions obtained from all the three mocks are shown in
Fig.\,\ref{fig:figure/chmf}. As one can see, the results for \texttt{Mock-2}
and \texttt{Mock-3} are very similar to that for \texttt{Mock-1}, except at
the massive end where the halo mass functions are overestimated because of the
larger uncertainty in the halo mass.

\section{The brightest central galaxies of clusters and their progenitors}%
\label{sec:local_brightest_central_galaxies_and_their_progenitor_galaxies}

\begin{figure*}
    \centering \includegraphics[width=0.8\linewidth]{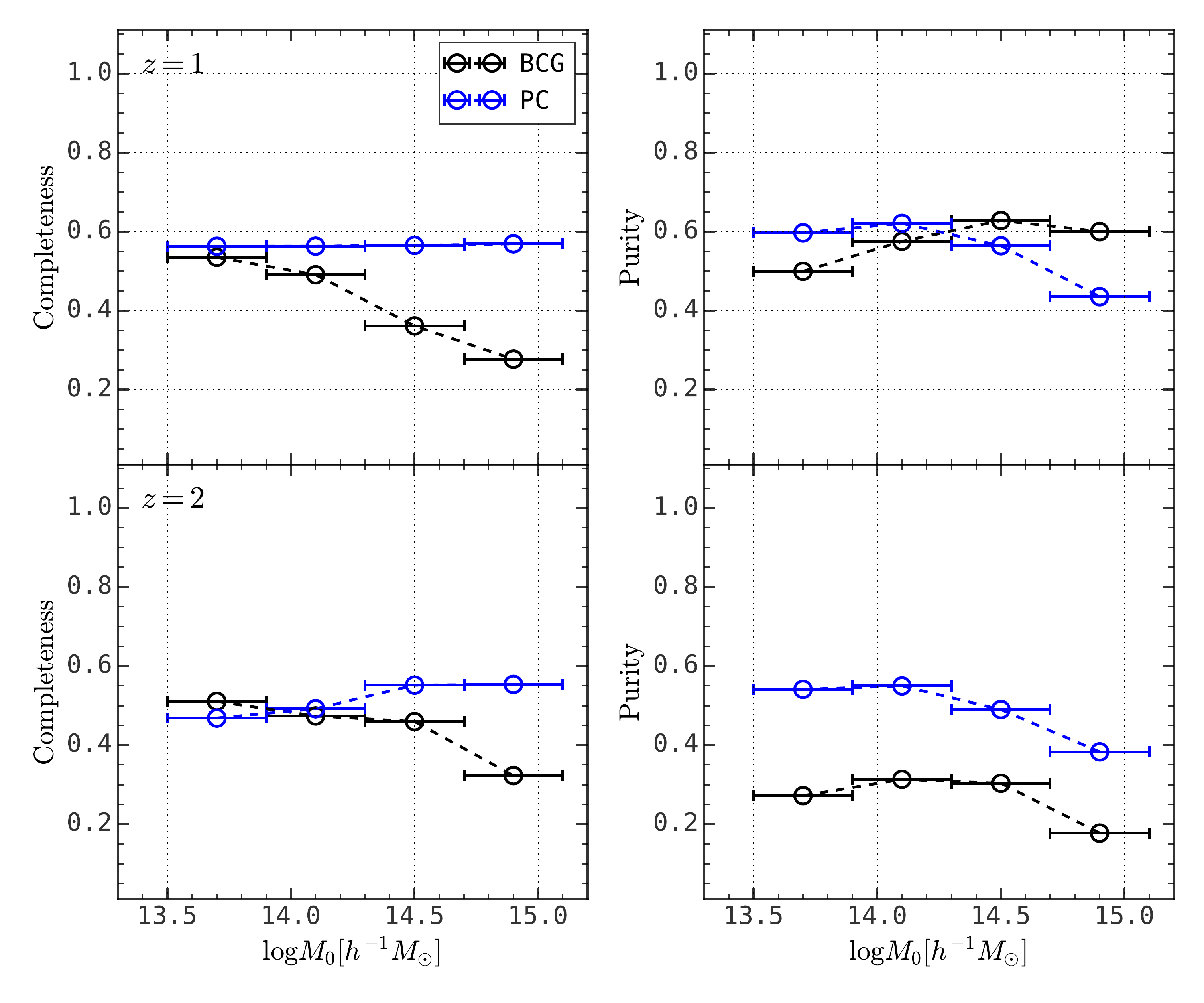}
    \caption{Comparison of the performance in connecting central galaxies at
    $z=0$ and $z=1~(2)$. The black lines are for \texttt{BCG} method and blue
    lines are for \texttt{PC} method (See text). The upper panels are for $z=1$
    and lower panels are for $z=2$.}%
    \label{fig:figure/pc_bcg_compare}
\end{figure*}

Some previous investigations have attempted to link the brightest cluster
galaxies across different redshifts by integrating the mass accretion rate to
get the descendant halo mass at $z=0$ using the formula in
\cite{fakhouriMergerRatesMass2010}
\citep[See][]{lidmanEvidenceSignificantGrowth2012, cookeStellarMassGrowth2019,
demaioGrowthBrightestCluster2020, linSTELLARMASSGROWTH2013}. Here we compare
our method with this, and we denote the method of
\cite{lidmanEvidenceSignificantGrowth2012} as the \texttt{BCG} method and ours
the \texttt{PC} method for convenience. For central galaxies at $z=0$ with halo
mass in a mass bin, the \texttt{BCG} method identifies all the halos at a given
redshift (in our case, $z=1$ or $z=2$) whose descendant mass at $z=0$ is in the
same mass bin, while the \texttt{PC} method selects all central galaxies of the
most massive halos (MMH) in candidate proto-clusters whose estimated descendant
halo mass is in a given mass bin. In each descendant halo mass bin, we define
the {\bf completeness} as the fraction of centrals at $z=0$ whose progenitors
are selected at $z=1$ or 2, and the {\bf purity} as the fraction of the selected
galaxies at $z=1$ or 2 which are the true progenitors of the central galaxies
in the descendant halo mass bin in question. The comparison is presented in
Fig. \ref{fig:figure/pc_bcg_compare}, with black lines showing results for the
\texttt{BCG} method and the blue lines for the \texttt{PC} method. At $z=1$,
the \texttt{BCG} method performs slightly better in purity at the massive end,
but it is at the cost of a much worse performance in completeness. At $z=2$,
the \texttt{PC} method performs better, especially in purity. Note that the
absolute values of the completeness and purity depend on the choice of the mass
bin size, because the mass bin size here is equivalent to the tolerance of
descendant halo mass error.

\section{Test using Illustris-TNG}%
\label{sec:abundance_matching_for_illustris_tng}

\begin{figure*}
    \centering
    \includegraphics[width=1\linewidth]{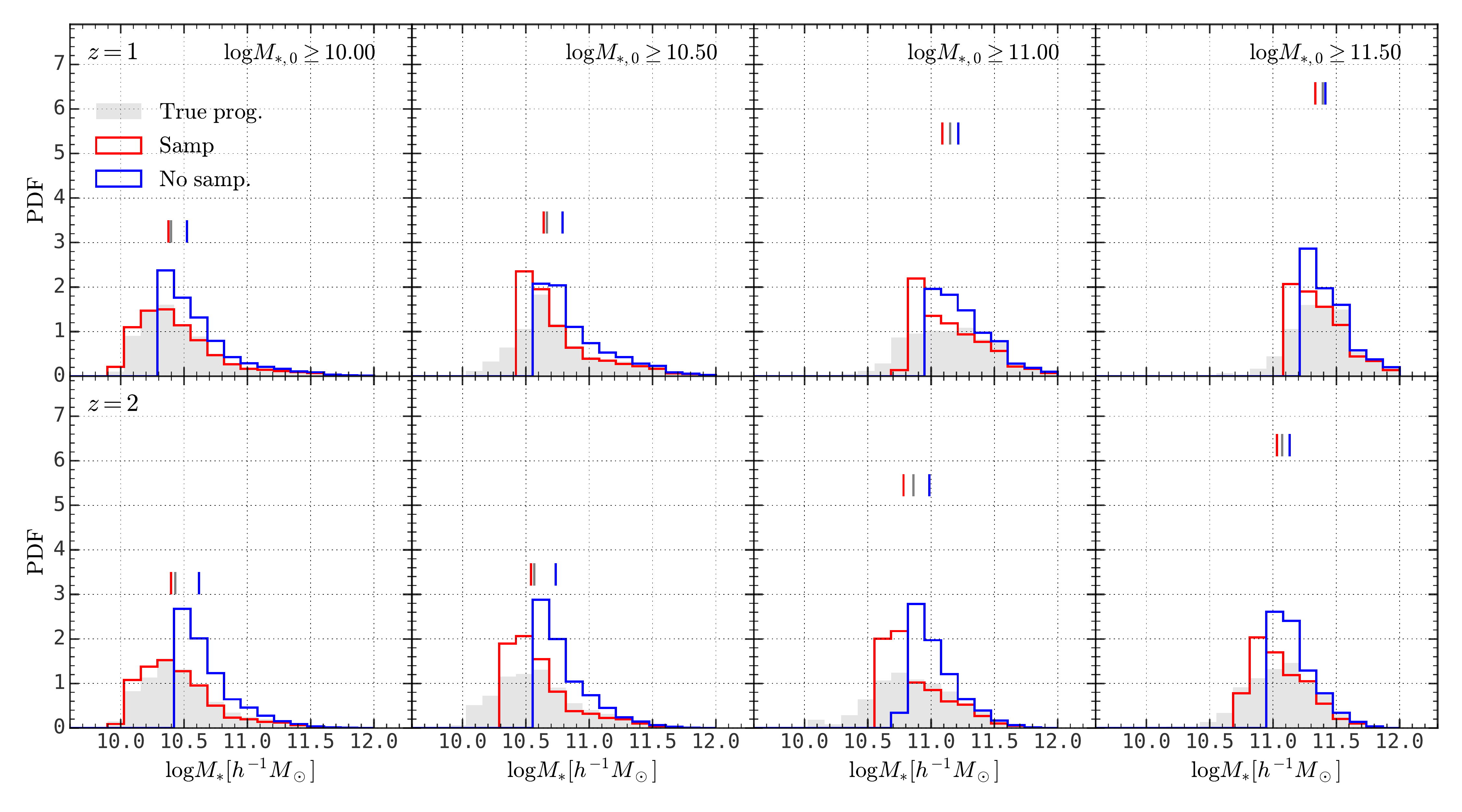}
    \caption{Same as Fig.\,\ref{fig:figure/galaxy_prog_hist_with_am}, just for galaxies
    in Illustris TNG300-1 simulation.}%
    \label{fig:figure/galaxy_prog_hist_with_am_tng}
\end{figure*}

As a test, we apply the method in \S\,\ref{ssub:abundance_matching} to galaxies
in Illustris TNG300-1 simulation
\citep{nelsonIllustrisTNGSimulationsPublic2019,
pillepichFirstResultsIllustrisTNG2018}. In
Fig.\,\ref{fig:figure/galaxy_prog_hist_with_am_tng}, we show the stellar mass
distribution of the most massive progenitors for galaxies at $z=0$ with
$M_0\geq 10^{14}h^{-1}M_\odot$. The gray histogram shows the true distribution,
where the most massive progenitors are identified from the galaxy merger tree
in TNG300-1. One can see that the true distribution is distinct from the
results shown in Fig.\,\ref{fig:figure/galaxy_prog_hist_with_am} due to the
difference in the galaxy formation models. To recover this distribution, we
first perform the same abundance matching as in
\S\,\ref{ssub:abundance_matching}, and the results are shown as the blue
histograms. The predicted distributions are biased towards the high mass end
due to mergers of galaxies. To deal with this problem, we again randomly select
galaxies in TNG300-1 at $z=1$ and $z=2$ according to the $F_{\rm mmp}$ shown in
Fig.\,\ref{fig:figure/mmp_probability}. We emphasize that the $F_{\rm mmp}$
model is adopted from the empirical model instead of the TNG300-1 simulation.
We then apply the abundance matching to find the most massive progenitors for
galaxies at $z=0$, and the results are shown as the red histograms. The bias
relative to the true distribution is now much reduced, indicating that $F_{\rm
mmp}$ is independent of the details of the galaxy formation model.

\bsp	
\label{lastpage}
\end{document}